\documentclass[preprint2,usenatbib]{mnras}
\usepackage{graphicx}
\usepackage {amsmath,amssymb}
\usepackage{color}
\usepackage{aas_macros}
\usepackage{array}
\title[On the environment of LSBs at different scales]{On the environment of Low Surface Brightness galaxies at different scales}

\author[P\'erez-Monta\~no \& Cervantes Sodi]{
Luis Enrique P\'erez-Monta\~no$^{1}$, %\thanks{E-mail: l.perez@irya.unam.mx}
Bernardo Cervantes Sodi$^{1}$%\thanks{E-mail: b.cervantes@irya.unam.mx}
\\
$^{1}$Instituto de Radioastronom\'ia y Astrof\'isica, UNAM, A.P. 3-72, C.P. 58089, Michoac\'an, M\'exico\\
}

\date{Accepted XXX. Received YYY; in original form ZZZ}

\pubyear{2019}

\begin{document}
\label{firstpage}
\pagerange{\pageref{firstpage}--\pageref{lastpage}}
\maketitle

\begin{abstract}

We select a volume-limited sample of galaxies derived from the SDSS-DR7 to study the environment of low surface brightness (LSB) galaxies at different scales, as well as several physical properties of the dark matter haloes where the LSB galaxies of the sample are embedded. To characterize the environment we make use of a number of publicly available value-added galaxy catalogues. We find a slight preference for LSB galaxies to be found in filaments instead of clusters, with their mean distance to the nearest filament typically larger than for high surface brightness (HSB) galaxies. The fraction of isolated central LSB galaxies is higher than the same fraction for HSB ones, and the density of their local environment lower. The stellar-to-halo mass ratio using four different estimates is up to $\sim$20\% for HSB galaxies. LSB central galaxies present more recent assembly times when compared with their  HSB counterparts. Regarding the $\lambda$ spin parameter, using six different proxies for its estimation, we find that LSB galaxies present systematically larger values of $\lambda$ than the HSB galaxy sample, and constructing a control sample with direct kinematic information drawn from ALFALFA, we confirm that the spin parameter of LSB galaxies is 1.6 to 2 times larger than the one estimated for their HSB counterparts.

\end{abstract}


%%%%%%%%%%%%%%%%%%%%%%%%%%%%%%%%%%%%%%%%%%%%%%%%%%

%%%%%%%%%%%%%%%%% BODY OF PAPER %%%%%%%%%%%%%%%%%%

\section{Introduction}

Galaxies as usually classified according to their morphological types given in therms of the classification scheme introduced by Hubble \citep{Hubble26}, in which galaxies are segregated into two main groups; ellipticals and spiral galaxies, with these further classified as barred and unbarred ones depending on the presence or not of stellar bars at their centres. Although this classification is popular and gives us a general idea of the physical properties of a given galaxy, it is by no means the only way to classify them. Taking in consideration their star formation activity, galaxies can be classified as star forming or quenched; focusing our attention in their nuclear activity, they can be AGN or non-AGN galaxies, and regarding their surface brightness they fall into two main groups, low or high surface brightness galaxies. Traditionally defined as galaxies with surface brightness lower than the dark night sky \citep{Impey97}, \citet{Freeman70} defined low surface brightness (LSB) galaxies as those having a central surface brightness in the $B-$band lower than
$\mu_B = 21.65$ mag arcsec $^{-2}$, and more recent works use limiting surface brightness magnitudes in the $r-$band \citep{Courteau96} or the $K_s-$band \citep{Monnier03} to distinguish between LSB and HSB galaxies.

LSB galaxies are an interesting subset of the galaxy population, characterized by their low disc stellar surface density and blue colours \citep{Vorobyov09}, these galaxies present low star formation rates \citep{vanderHulst93,vanZee97,Wyder09,vanderHoek00,Schombert11}, high gas fraction and total HI masses \citep{Burkholder01,Oneil04,Huang14,Du15}, low dust content \citep{Hinz07,Rahman07} and low metallicities \citep{deBlok98,deBlok98_b,deNaray04}, which suggest that LSB galaxies are less evolved that their HSB counterparts, indicating that they follow different evolutionary paths or at least they evolve in a much lower rate.

A common scenario proposed for the formation of LSB galaxies is that in which they are formed in dark matter haloes with high angular momentum. If baryons and dark matter share the same specific angular momentum \citep{Fall80}, then the disc scale-length is regulated by the dark matter halo spin \citep{Hdz98,Mo98,KimLee13} and the low surface brightness of the disc, which is a direct consequence of the low stellar mass density, is set by the value of the $\mathrm{\lambda}$ spin parameter of the whole configuration; the higher the spin parameter, the lower the stellar surface mass density. In this sense, LSB galaxies would be systems from the high-spin tail of the total galaxy distribution \citep{Dalcanton97,Jimenez98,Mo98,Boissier03,Jimenez03,KimLee13}. 

If LSB galaxies are indeed formed in high spinning dark matter haloes, then their sparse stellar discs are subdominant at all radii, and the dark matter halo, which is the dynamically dominant component \citep{Pickering97,McGaugh01}, help to stabilize the disc \citep{Ostriker73,DeBuhr12,Yurin15,Algorry17}. Moreover, given that not only the stellar discs are sub-dominant, but also embedded in high spinning haloes, according to various studies \citep{Mayer04,Long14,Fujii19}, the formation of stellar bars in LSB galaxies should be suppressed, and when formed, their growth would be quenched forming only short size, weak bars. The low fraction of barred LSB galaxies is a strong evidence in support of this scenario \citep{Cerv13,Honey16,Cerv17}.

At larger scales, beyond the one halo term, using a sample of $\sim$ 340 LSB disc galaxies, \cite{Bothun93} found that the average distance  between the target galaxies and their closest neighbour was about 1.7 times larger than the same statistic for HSB galaxies, suggesting that LSB are formed in relative isolation. This result received validation with the work by \cite{Mo94}, who reports a lower amplitude of the correlation function for LSB galaxies than for the general sample employed in their study (CfA and \textit{IRAS}), indicating that they are less clustered. Using the early release of the SDSS, \citet{Rosenbaum04} show differences in the local environmental density on scales from 2 to 5 Mpc between LSB and HSB galaxies, supporting the idea of gas-rich LSB galaxies forming in low density regions without frequent galaxy interactions, a result later confirmed in \citet{Rosenbaum09} using a larger sample of galaxies, where introducing an anti-bias parameter the authors sustain that LSB galaxies form in low-density regions and then drift to the outer part of filaments and walls of the large-scale that conforms the structure of the Universe.

\citet{Galaz11} report a lack of companions for LSB galaxies when compared to HSB ones at scales $<$2 Mpc, with an increase of galaxy interactions form LSB galaxies with strong star formation activity, suggesting than rather than being formed in low density environments, the isolation of these kind of galaxies is a condition for their survival. And, for the case of bulgeless LSB galaxies, \cite{Shao15} report similar stellar populations in low and high density regions, suggesting that the environment may play only a secondary role in their evolution, while mergers are identified as a major role in the evolution of the more massive systems.

In the present work we study the environment of LSB galaxies at different scales, from the dark matter halo embedding the galaxy, characterized by its mass, spin and assembly time,  to the local density at different smoothing scales and the large-scale structures these galaxies inhabit, in order to explore at which scale we find a larger difference between LSB and HSB galaxies. To account for the halo mass, spin and halo assembly time, we will adopt observational proxies to explore if LSB galaxies are indeed formed in high spinning haloes, as predicted by theoretical works, determine if the halo mass proxies provide lower stellar-to-halo mass ratios for LSB galaxies and determine if their assembly times are more recent as an indicator of their different evolutionary rate. Regarding the environment at larger scales we aim to confirm previous findings that LSB galaxies are located in low-density regions and explore if they are present in the same large-scale structures where HSB galaxies can be found.

The paper is organized as follows: in Section \ref{sec:Sample} we describe the Value-Added Galaxy Catalogues used to construct the volume-limited and a control samples used throughout the paper. In Section \ref{sec:DMhalo} we describe the dark matter halo mass estimates and the proxy for the halo assembly time. Section \ref{sec:SpinCalculation} presents the method to estimate the angular momentum and spin of the systems, while Section \ref{sec:Results} contains the main results with a summary and conclusions of the work presented in Section \ref{sec:Conclusions}. Throughout this paper, we use a cosmology with density parameter $\Omega_{\mathrm{m}}$ = 0.3, cosmological constant $\Omega_{\mathrm{\Lambda}}$ = 0.7 and Hubble-Lemaitre constant written as  $H_{0}  = 100$ $h$ km s$^{-1}$ Mpc$^{-1}$, with $h=0.7$.

\section{Sample Selection}
\label{sec:Sample}

In this work we make use of a number of public catalogues to segregate the galaxies in our sample into LSBs and HSBs, to give different estimates of the specific angular momentum and to characterize their environment at different scales. In what follows we briefly describe each of them. 

\subsection{SDSS optical data}

We build our galaxy sample using the KIAS Value-Added Catalogue \citep[KIAS VAC]{Choi10}, which is drawn from the Sloan Digital Sky Survey Data Release 7 \citep[SDSS DR7]{Abazajian09}, as our base catalogue. From this catalogue we retrieve photometric information such as the corrected exponential magnitude in the \textit{g, r} bands, absolute $r-$band magnitude $M_r$, and the axis ratio $q=b/a$. We also took directly from SDSS DR7 website the different exponential radii $\alpha$ in the above-mentioned bands in order to estimate the central surface brightness in the $B-$band, to segregate LSBs from HSBs. Total stellar masses $M_{*}$ and star formation rates (SFR) are extracted from the MPA/JHU SDSS database \citep{Kauffmann03,Brinchmann04}\footnote{\url{http://www.mpa-garching.mpg.de/SDSS/}}. Finally, we include structural information from the \citet{Simard11} galaxy catalogue, where two-dimensional bulge+disc decompositions in the $g$ and $r$ bands are provided for a large sample of $\sim$ 1,000,000 galaxies from the SDSS DR7. We made use of the catalogue with fixed Sérsic index $n_{b}=4$ for the bulge component.

We incorporated the use of several catalogues to characterize the environment at different scales. To start, we made use of \citet{Yang07} group catalogue, where galaxy groups are identified by the method developed by \citet{Yang05}. In \citet{Yang07}, they compute the stellar mass and the characteristic luminosity of a given group, following an iterative process assuming a constant mass-to-light ratio, to obtain the mass of the dark matter halo where the group resides. This information allows us to separate our sample between centrals and satellites, where in our case the central corresponds to the most massive galaxy in the group. 

We also incorporate the analysis of the cosmic density field conducted by  \citet{Jasche10}, using a non-linear, non-Gaussian and fully Bayesian sample algorithm, which enables the inference of the highly non-linear density field for the SDSS DR7, with a grid resolution of $\backsim$ 3 Mpc, corrected by the survey geometry and the selection function of the SDSS. As a result, filaments, voids and cluster are clearly identified and a probability is assigned to each galaxy to belong to the different structures. We assume that a given galaxy belongs to a given structure if this probability is $\geq$ 0.5. A measure of the local overdensity $\delta=(\rho - \bar{\rho})/\bar{\rho}$ with a smoothed scale of 3 Mpc, is also provided.

Another way to characterize the environment in which galaxies reside, is to look at their distance to their nearest filament axis ($d_{fil}$), information available from the \citet{Tempel14a} catalogue, which applies an object point process with interactions to trace and extract the filamentary network from the SDSS data. The method relies on the assumption that galaxies assemble randomly into cylinders having similar orientations. Two or more cylinders, connected and aligned form a filament. The probabilistic nature of the method provides  the likelihood of the detected filaments, together with the filament orientation field, using a Bayesian framework. With the radius of the filaments set to 0.5 $h^{-1}$ Mpc, the galaxy population can be segregated into galaxies inside and outside filaments. The chosen radius of the filaments is set to be similar to the scale of galaxy groups/clusters, where the strongest impact for galaxy formation and evolution is expected. 

Finally, we included a $r-$band luminosity density smoothed over different scales ($a=$ 1, 2, 4 and 8 $h{-1}$ Mpc) with a B$_3$ spline kernel, from \cite{Tempel14b}, where a correction for galaxies that lie outside of the observational window of the survey, is implemented, based on the luminosity limits of the observational window. This catalogue also provides the probability of a galaxy to be an S0, Sab or Scd, taken directly from \citet{HuertasCompany11}.

\subsection{ALFALFA HI data}

In order to obtain direct measurements of $V_{rot}$, we use the Arecibo Legacy Fast ALFA survey $\alpha$.100 \citep{Haynes18} \footnote{\url{http://egg.astro.cornell.edu/alfalfa/data/index.php}}, which is a blind, single-dish, flux-limited extragalactic HI survey, conducted using the 305 meter Arecibo telescope designed to sample the HI mass function, over a distance of $\backsimeq$ 100 Mpc. The completed survey contains about 31,500 extragalactic HI line sources, over a wide solid angle of $\backsim$ 7000 deg$^2$ \citep{Giovanelli05, Giovanelli&Haynes15}. The full description of the catalogue is available on \citet{Haynes18}. The $\alpha$.100 catalogue includes detections of HI sources with  the corresponding velocity width of the HI line profile, W$_{50}$, measured at the 50$\%$ level of each of the two peaks, and corrected for instrumental. We compute the galactic rotational velocity as
\begin{equation}
	\label{vrot}
 V_{rot}=\frac{W_{50}}{2 \times \sin i},	
\end{equation}

where $i$ is the disc inclination relative to the sky plane, as provided by \cite{Simard11} (assuming the same orientation for the gas and the stellar discs).

\subsection{Central Surface Brightness} 

In order to segregate the galaxies in our main sample into LSBs and HSBs, we follow \cite{Trach06} and \cite{Zhong08} to estimate the central surface brightness as
\begin{equation}
	\label{mu}
  \mu_X=m_{X} + 2.5 \log{(2\pi \alpha^2 q)} -10 \log{(1+z)},
\end{equation}

  where $m_{X}$ is the apparent magnitude measured in any $X$-band, $\alpha$ the exponential radius, $q$ the axis ratio of the disc, and $z$ the redshift. This expression is corrected by inclination and cosmological dimming, and applied to both $g-$ and $r-$ bands, to finally obtain the central surface brightness in the $B-$band using the transformation equation by \citet{Smith02} 
\begin{equation}
\label{muB}
  \mu_B = \mu_g + 0.47 (\mu_g - \mu_r) + 0.17.		
\end{equation}  

We are now able to define our sub-sample of LSBs as those with $\mu_0(B) \geq 22.0$ mag arcsec$^{-2}$, consisting of 132,154  galaxies (25\%) with the rest 404,582  (75\%) being HSB galaxies. This main sample would be the starting point to build our volume-limited and our control samples, as described in the next subsections.

\subsection{Volume-Limited Sample} 

\begin{figure*}
\centering
 \includegraphics[width=0.8\textwidth]{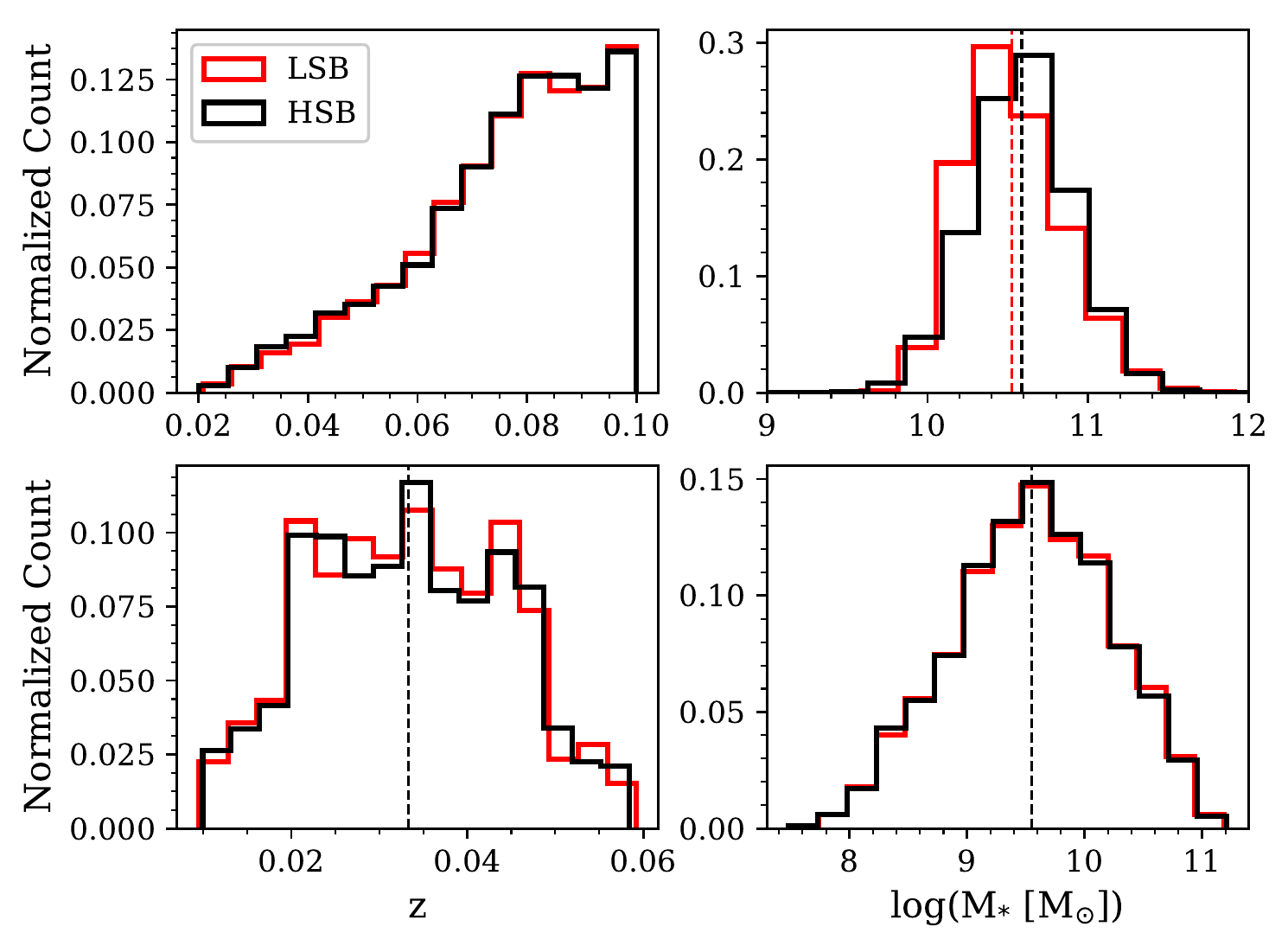}
\caption{\textit{Upper panels:} Redshift (left) and stellar mass (right) distributions for LSB (red lines) and HSB (black lines) galaxies of the VL sample. \textit{Lower panels:} Same distributions, but for the galaxies on the CS. For both, dotted lines represent the mean value of the distribution. The colour convention to distinguish LSB and HSB galaxies will be kept thought the paper.} 
\label{fig:CS}
\end{figure*}

  In order to avoid completeness problems, and to drew results that are directly comparable with other studies focused on LSB selected using SDSS data \citep{Zhong08, Galaz11}, we construct a volume-limited sample (hereafter, VL) with galaxies brighter than $ M_r=-19.8$ mag within a redshift range of $0.01 < z < 0.1$. We restrict our sample to galaxies nearly face-on to avoid extinction correction following, the criteria $ q > 0.4 $. Besides, since our interest is centred on spiral galaxies, we also select those galaxies with $fracDev < 0.9 $, where $fracDev$, as provided by the SDSS database, accounts for the fraction of total flux fitted by a de Vaucouleurs profile, in order to include only late-type galaxies. The final sample is conformed by 64,351 galaxies, of which  21,273 are LSBs and 43,078 HSBs. $z$ and $M_{*}$ distributions for the VL sample are shown in the upper panels of Fig. \ref{fig:CS}.

\subsection{Control Samples}

When estimating the specific angular momentum and spin of the galaxies in our sample, we make use of the match between our main optical sample and the ALFALFA $\alpha$.100 catalogue, including galaxies with inclination angle $i$ in the range 25$^{\circ} < i < $75$^{\circ}$. This guaranties a correct measurement of the disc rotation velocity avoiding face-on systems, and avoid strong extinction for nearly edge-on galaxies. Unfortunately, this reduces the number of galaxies in our sample to only $\sim$ 8,000 sources, and hinders the compilation of a volume-limited sample. In order to compare the physical properties of LSB and HSB galaxies, we build control samples (hereafter CS) for LSB and HSB galaxies, by randomly selecting galaxies from the sample of HSB galaxies having the same distribution of stellar mass and redshift as the LSB sample, as shown in Fig. \ref{fig:CS}. The resulting control samples consist of 2,473 galaxies each one. We verify that we cannot reject the null hypothesis that the stellar mass and redshift distributions of both samples are extracted from the same parent population by performing a Kolmogorov-Smirnov (KS) test, with corresponding p-values of  0.988 and 1 respectively.

\section{Dark matter halo mass and assembly time}
\label{sec:DMhalo}
One way to study the environment of LSB galaxies is to look at the mass of the hosting haloes. Here we list our five halo mass estimates that will also be used in the next section to estimate the spin parameter of the galaxies in the different samples previously described, as well as to give a proxy for the halo assembly time.\\

1. For our first estimate we adopt a constant value for the stellar-to-halo mass ratio of 1/25, using the Milky Way (MW) as a representative example \citep{Wilkinson99, HernandezAvila01}. Although this choice is over-simplistic, we include it because it is the one employed by \cite{HdzCerv06} to estimate the $\lambda$ parameter, as described in the next section, and has been extensively used by various authors \citep{Berta08, Gogarten10, Huang12, Wang18}. Choosing a different stellar-to-halo mass ratio as the one presented by recent studies \citep{Battaglia05, McMillan11, Kafle14} would only shift the distributions presented in the top left panels of Figures \ref{VolLimSpins} and \ref{ContSampSpins}, without changing qualitatively our results.\\

2. Motivated by observational studies \citep{Zavala03, Pizagno05}, where the ratio of halo-to-stellar mass correlates with disc surface density, \cite{Gnedin07} proposed a model where this fraction depends on the stellar surface density according to: 
\begin{equation}
	\label{Gnedin}
f_{*}=f_0 \left(\displaystyle \frac{M_{*}R_d^{-2}}{10^{9.2}M_{\odot}km \ pc^{-2}} \right)^{p},
\end{equation}
with $p=0.2$, and $f_0$ is chosen using the Milky Way as a representative example. \\

3. \cite{Pap11} studied how galaxies of different circular velocities populate dark matter haloes through an abundance matching procedure that assumes the existence of a one-to-one relationship between the dark matter halo and galaxy circular velocities, based on the premise that the space density of haloes with circular velocity larger than a given velocity $V$, should be the same to the density of galaxies with rotational velocities larger than the value imposed by $n(V_{h} > V)= n(V_{rot} > f(V))$. From $V_{h}$, we estimate the virial mass of the halo using the tight correlation between $V_{h}$ and $M_{h}$ found by \cite{Klypin11} using the Bolshoi simulation:
\begin{equation}
	\label{Klypin}
	V_{h}(M_{H})=2.8\times 10^{-2} (h M_{H})^{0.36}	.
\end{equation}

4. Given that the \cite{Pap11} relation is less constrained at the high mass end, it does not present the trend of decreasing $f_{*}$ for increasing stellar mass, above $M_{*} \sim$10$^{10.5} M_\odot$. To test if our results regarding the stellar-to-halo mass fraction and spin distribution of LSBs and HSBs changes with a halo mass estimate that presents this turn, we adopt the double power law derived by \cite{Hudson15} using galaxy-galaxy weak lensing, that directly provides the halo mass as a function of stellar mass through 
\begin{equation}
	\label{Hudson}
	f_{*}=2f_1 \left [\displaystyle \left (\displaystyle \frac{M_{*}}{M_1} \right )^{-0.43} + \left (\displaystyle \frac{M_{*}}{M_1} \right ) \right ]^{-1},
\end{equation}

where $f_1(z)=0.0357 + 0.026 (z-0.5)$ and $\log_{10}(M_1)=11.04 + 0.56 (z-0.5)$.\\

5.  The \citet{Yang07} galaxy group catalogue provides halo masses for each of the identified groups above a certain threshold mass, applying a stellar mass ranking method, from which the most massive galaxy of each group is defined as central, while the rest are treated as satellite galaxies. The groups of the catalogue are identified using a friends-of-friends algorithm, developed by \cite{Yang05}, with a linking length tuned with a dark matter halo simulation. For the central galaxies of each group, the stellar-to-halo mass ratio is directly the ratio between its stellar mass and the halo mass of the group.

  A quantity that can be derived from the galaxy group catalogue is the halo assembly time, defined to be redshift at which the main progenitor of the halo gathered half of its final mass. Since the halo assembly time is not directly observable, \cite{Wang11} shows that the halo formation time presents a tight correlation with the sub-structure fraction, $f_{s} = 1 - (M_{main}/M_{h})$, where $M_{main}$ is the mass of the main sub-halo at the centre of the host halo. Given that the mass of the host halo is provided in the group catalogue and that $M_{main}$ can be estimated using sub-halo abundance matching, \cite{Lim16} define the quantity 
\begin{equation}
	\label{HAT}
		f_c\equiv\frac{M_{*,c}}{M_h},
	\end{equation}
as a proxy of the halo assembly time $z_{f}$. Here, $M_{*,c}$ is the stellar mass of the central galaxy of the group, and $M_{h}$ its halo mass. For the case of central galaxies, this proxy corresponds to the natural stellar-to-halo mass ratio, for satellites $f_c$ is that of the central galaxy of the group.

\section{Angular momentum and spin parameter}
\label{sec:SpinCalculation}

In order to give an estimate of the spin parameter, we start by calculating the specific angular momentum of the galaxies in our sample. Given that we count with bulge+disc decompositions from the \cite{Simard11} catalogue, the total specific angular momentum will be given by
\begin{equation}
	\label{lstars}
	j_{*}=f_b j_b + (1-f_b) j_d,		
\end{equation}

where $f_b$ is the $r-$band luminosity bulge fraction from \citet{Simard11}, and $j_b$ and $j_d$ are the specific angular momentum of the bulge and disc components respectively. For an infinitely thin disc with an exponential surface density profile and a rigorously flat rotation curve, the specific angular momentum is given by 
\begin{equation}
\label{jdisk}
  j_d=2 V_{rot} R_d,
\end{equation}
with $R_d$ the disc scale-length taken in the $r-$band and $V_{rot}$ is estimated using equation \ref{vrot} for the control samples. For the case of the galaxies in the volume limited sample, $V_{rot}$ is assigned adopting the stellar-mass Tully Fisher relation from \citet{Reyes11}.

Following \citet{RomFall12}, we estimate the specific angular momentum of the bulge component with a de Vaucouleurs density profile in therms of the observed rotation velocity at some arbitrary location $v_s$, and the effective radius $R_e$;
\begin{equation}
	\label{jbulge}
  j_b=3.03 v_{s} R_e.
\end{equation}
$v_s$ is estimated indirectly, based on the ellipticity and the central velocity dispersion $\sigma_0$ through:
\begin{equation*}
 v_s= (v/\sigma)^{*} \sigma_0 \left(\displaystyle \frac{\varepsilon}{1-\varepsilon} \right)^{1/2},
\end{equation*}

where $(v/\sigma)^{*}$ describes the relative dynamical importance of rotation and pressure, which is fixed at $\sim$0.7 for spiral bulges that are near oblate-isotropic \citep{Binney&Tremaine08}.

To estimate the spin parameter, we adopt the model by \cite{HdzCerv06}, where the dark matter halo, with an isothermal density profile, is responsible for establishing a rigorously flat rotation curve throughout the entire galaxy. By assuming that the total potential energy of the galaxy is dominated by that of the halo, and that the entire system is virialized, we can replace $E$ in the expression for $\lambda$ by half the potential energy of the halo, given by $W=V_{rot}^{2}M_H$. In this way, the spin parameter is given by 
\begin{equation}
	\label{Spin}
  \lambda=\frac{j V_{rot}}{\sqrt{2}G M} 	,
\end{equation}

with $M$ the dynamical mass of the galaxy and $j$ the total specific angular momentum. Hereafter, eq. \ref{Spin} will be our general expression to calculate $\lambda$, particularly, for the case of $\lambda_{1,2,3,4,5}$ with the corresponding estimation for $M$ given in the previous section. For the specific angular momentum we assume that the specific angular momenta of disc and halo are equal, as generally assumed in galactic formation models  \citep{Fall80,Mo98}. This assumption will set our $\lambda$ estimates as upper limit values, in the case of angular momentum dissipation by the baryonic component.

Our first estimate ($\lambda_1$) considers only a disc component for the stellar distribution, a fixed stellar-to-halo mass ratio of $1/25$ and a baryonic Tully-Fisher relation \citep{Gurovich04} that allow us to recover the expression by \cite{HdzCerv06}:
\begin{equation}
	\label{SpinHdz}
	\lambda_1=21.8\frac{R_d}{(V_{rot})^{3/2}}	.
\end{equation}

Estimates $\lambda_{2,3,4,5}$, all take into account a bulge+disc decomposition and their corresponding stellar-to-halo mass estimates enumerated in Section \ref{sec:DMhalo}.

 Finally we compute $\lambda_6$ with the expression provided by \citet{Meurer18} where the spin can be written as a function of the orbital time $t_{orb}(R)=\frac{2\pi R}{v_{rot}}$, as
\begin{equation}
	\label{SpinClocks}
    \lambda_6=\frac{\sqrt{50}}{\pi}\frac{t_{orb}(R)}{t_H}\frac{R_d}{R},
\end{equation}
where $t_H$ is the Hubble time, and $R_d$ is the disc scale length. This expression is obtained from the study of HI galaxies, which obey linear relationships between their maximum radius and rotational velocity, assuming the same model of a disc galaxy dominated by an isothermal halo with a flat rotation curve.

\section{Results}
\label{sec:Results}

\subsection{Large scale environment and local density.}

Starting with our VL sample, we estimate the percentage of LSB and HSB galaxies inhabiting different large-scale structures (voids, sheets, filaments and clusters), according to the classification by \citet{Jasche10}. If the probability of a given galaxy to belong to a specific structure is larger than 0.5, that galaxy is identified as member of the structure in consideration. Table \ref{table:Jasche} presents the percentage of galaxies of each sub sample belonging to the different structures, showing no differences for the case of void or sheet galaxies, and only marginal differences for filament and cluster ones; there is a larger fraction of LSBs residing in filaments when compared with HSBs, while the opposite is found for cluster galaxies. Although the differences are minor, the distributions are statistically different as accounted by a KS test.

\begin{table}
	\centering
	\begin{tabular}{c c c} % centered columns
	\hline\hline 
	Large-Scale structure & LSB & HSB\\ [0.5ex] % inserts table
	%heading
	\hline % inserts single horizontal line
	Void 		 & $<$ 0.01\% &  0.01\% \\ % inserting body of the table
	Sheet 	 & 	16\% 		&  16\%   \\
	Filament &  57\% 		&  55\%  \\
	Cluster	 &  13\% 		&  15\%  \\ 
	\hline % inserts single horizontal line
	\end{tabular}
	\caption{Membership to large-scale structures.}
	\label{table:Jasche} 
\end{table}

We also look at the distance to the nearest identified filament $d_{fil}$ from the \citep{Tempel14a} catalogue, discriminating between LSB and HSB galaxies. The median distance to the nearest filament is larger for LSBs than for HSBs, as shown in Table  \ref{table:Tempel}, although this difference is only marginal as indicated by the KS test, yielding a $p-$value of 0.044.

\begin{figure}
\centering
\includegraphics[width=0.4\textwidth]{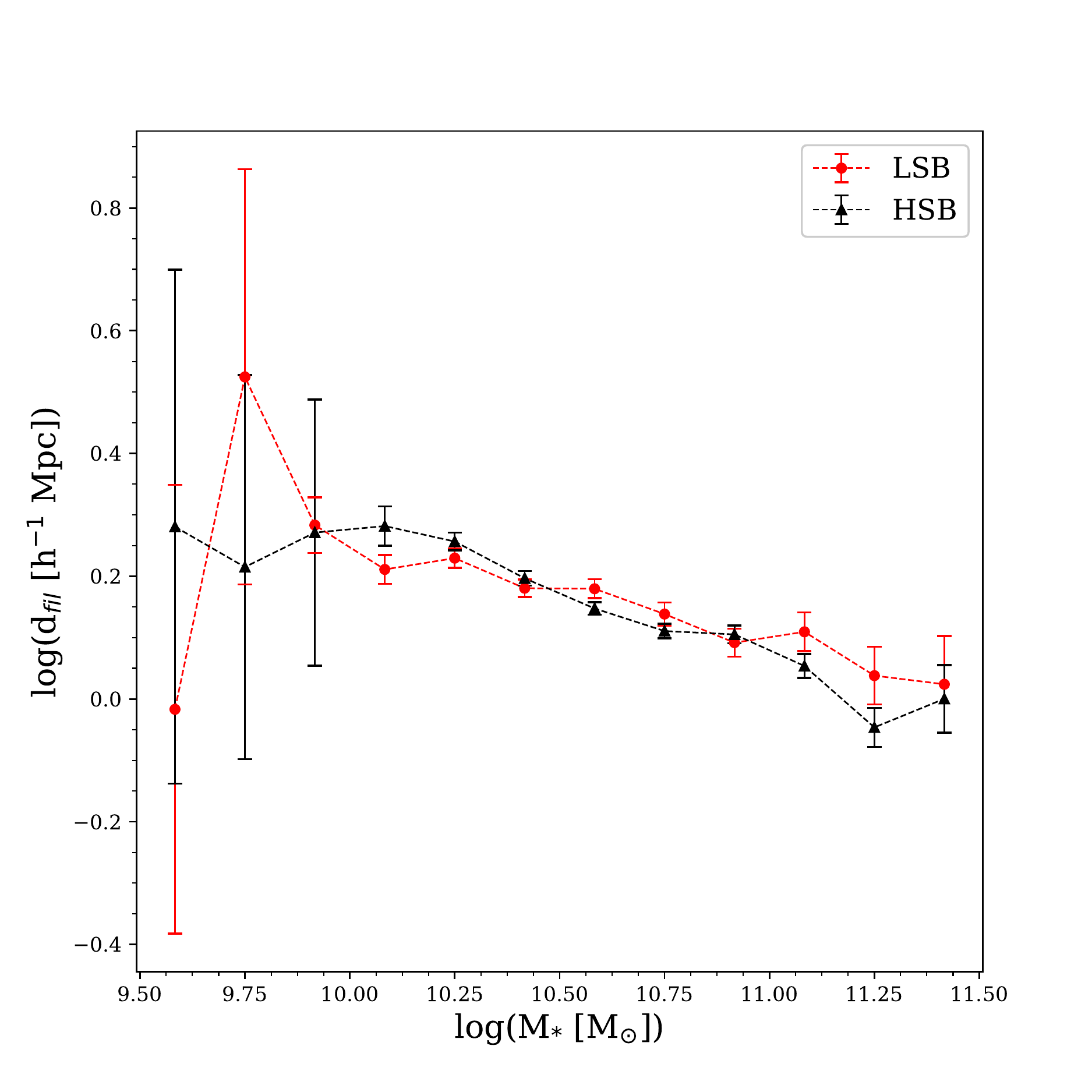}
\caption{Median value of the distance to the nearest filament $d_{fill}$, as a function of stellar mass M$_*$ for the full VL sample. Error bars denote the 1$\sigma$ confidence intervals computed using a bootstrap re-sampling method.}
\label{fig:dfill}
\end{figure}

\begin{figure}
\centering
\begin{tabular}{c}
\includegraphics[width=0.4\textwidth]{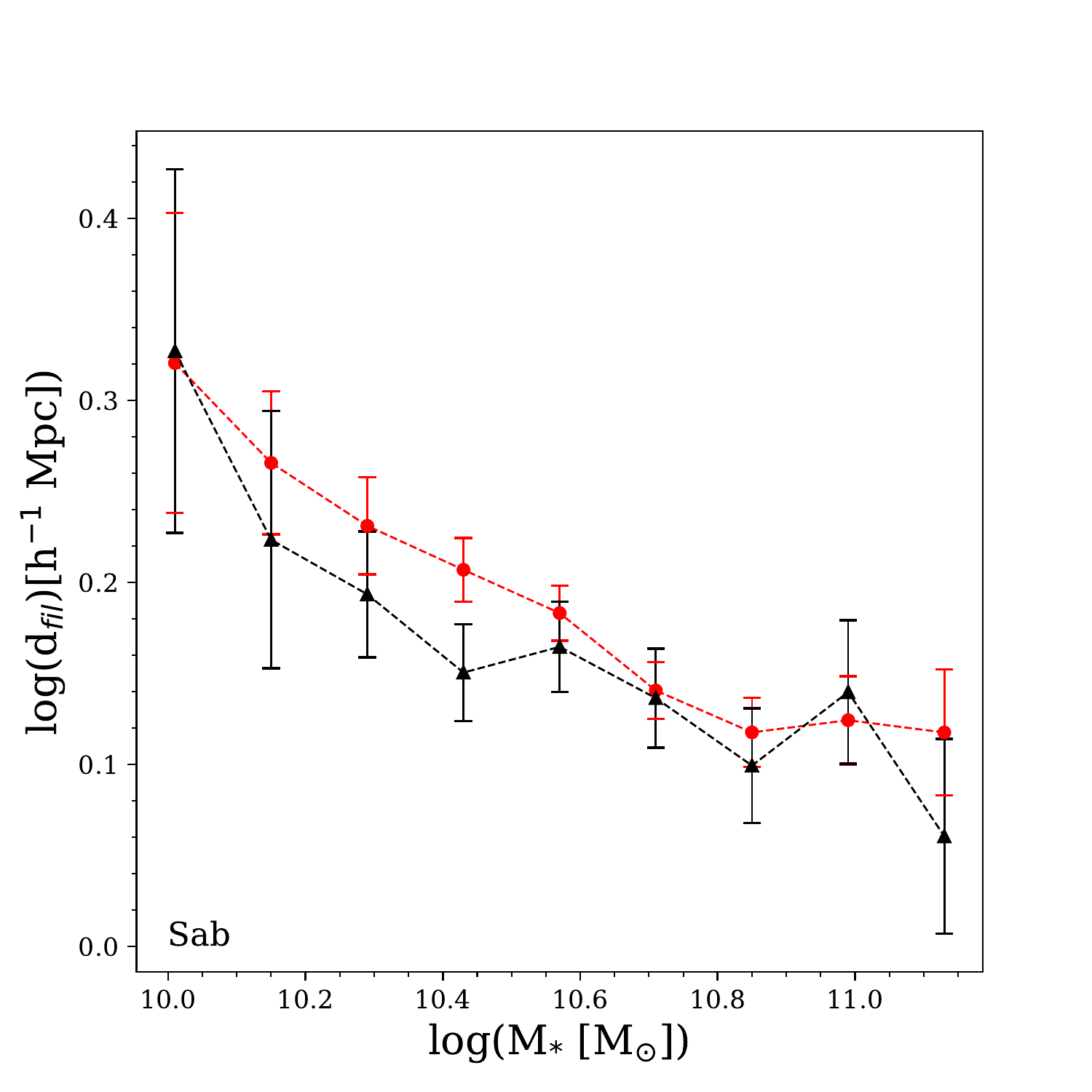}  \\ 
 \includegraphics[width=0.4\textwidth]{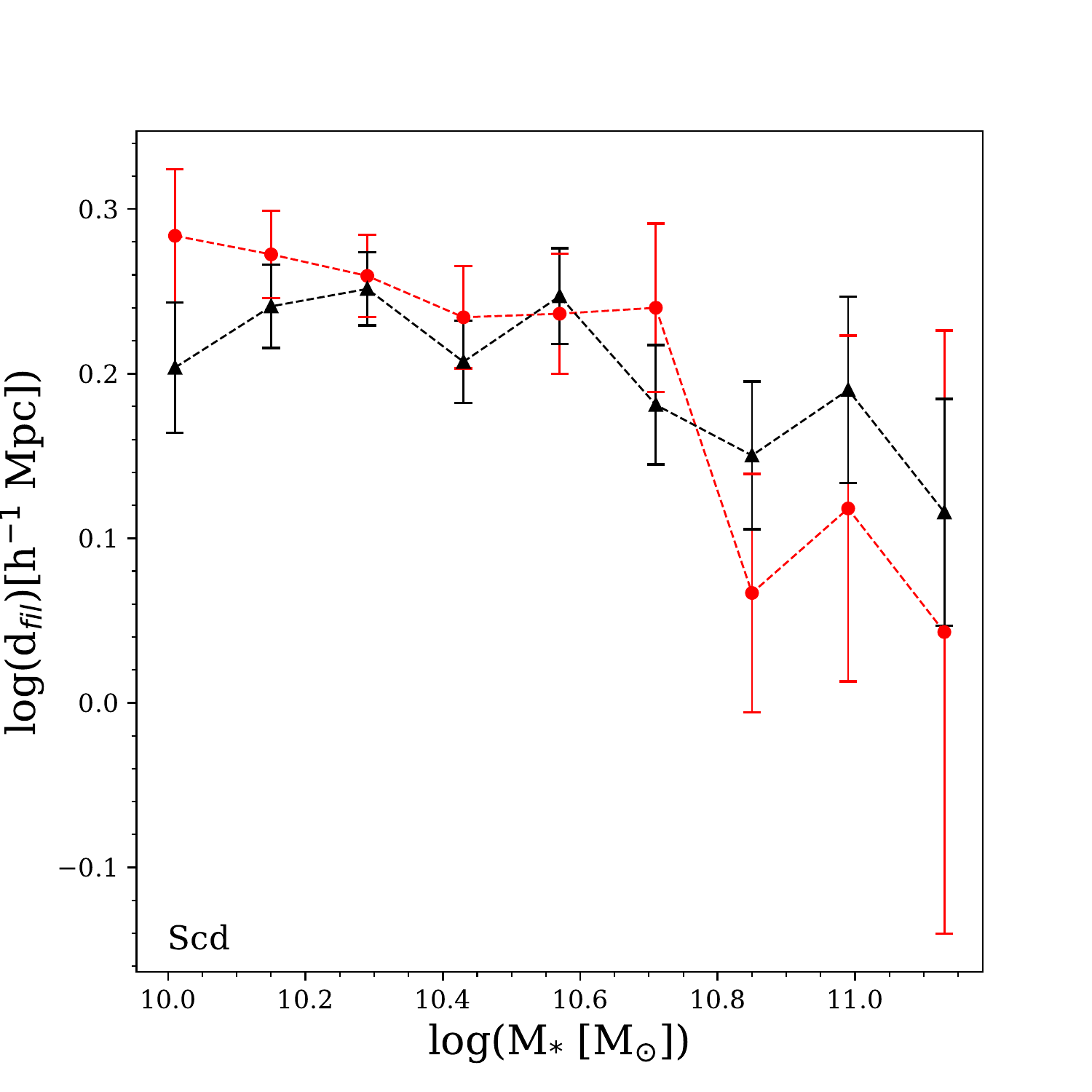}
\end{tabular}
\caption{Median value of the distance to the nearest filament $d_{fill}$ as a function of stellar mass M$_*$, segregating the VL sample between early- (top) and late-types (bottom).} 
\label{fig:morph-dfill} 
\end{figure}

Given that more massive galaxies tend to reside closer to the center of filaments \citep{Alpaslan16, Beygu17, Kuutma17}, in Figure \ref{fig:dfill} we show the median value of $d_{fil}$ as a function of stellar mass for the galaxies in the VL sample, in order to remove any mass dependence. Error bars are computed using a bootstrap re-sampling method and denote the estimated 1$\sigma$ confidence intervals based on a thousand random realizations derived from the original sample.  This method is used to assign error bars to all the results presented in this study. Within error bars, we find no difference between low and high surface brightness galaxies
however, when we segregate by morphological type (Fig. \ref{fig:morph-dfill}) we find, specially for early-type galaxies, that LSB are farther away from the nearest filament than HSB. This difference might arise if LSBs are formed in low-density regions (e.g. voids) and then these systems migrate to the outer regions of the filaments, where they are observed now \citep{Rosenbaum09}. These results are replicated if we consider only galaxies identified as belonging to filaments according to the \cite{Jasche10} catalogue.

\begin{table*}
	\centering % used for centering table
	\begin{tabular}{c c c c c c c} % centered columns
	\hline\hline 
	Type & $d_{fil}$ &  den($a=1$)  & den($a=2$) & den($a=4$) & den($a=8$) & ln (2+$\delta$) \\
	 -	 &	 h$^{-1}$ Mpc	&  h${^3}$ Mpc$^{-3}$ &  h${^3}$ Mpc$^{-3}$ & h${^3}$ Mpc$^{-3}$ & h${^3}$ Mpc$^{-3}$ & - \\ [0.5ex] % inserts table
	%heading
	\hline % inserts single horizontal line
	LSB & 1.489 & 44.755 & 10.353 & 3.824 & 1.456 & 1.423 \\ % inserting body of the table
	HSB & 1.423 & 46.896 & 10.644 & 3.892 & 2.002 & 1.478 \\
	\hline
	$p$-value & 0.044 & $<$0.001 & $<$0.001 & 0.001 & 0.012 & 0.053 \\ 
	\hline
	\end{tabular}
	\caption{Median values of the environmental properties for the VL sample.}
	\label{table:Tempel} 
\end{table*}

\begin{figure}
\centering
\includegraphics[width=0.4\textwidth]{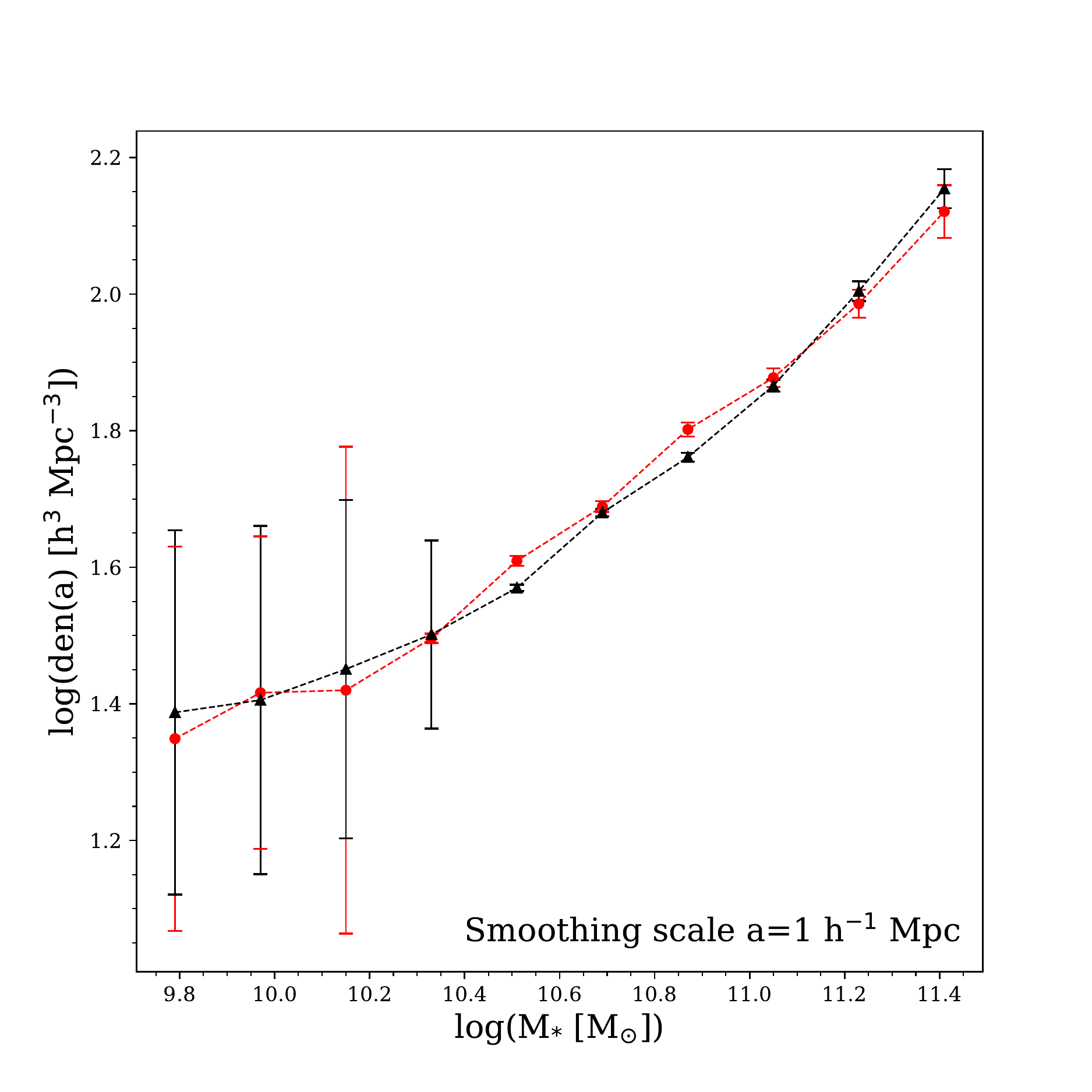}
\caption{Median value of the local density den($a$) as a function of stellar mass M$_*$, for the full VL sample. The label $a=1$ indicates that that the density is smoothed on 1 $h^{-1}$ Mpc scales.}
\label{fig:dena} %CHECK!!!
\end{figure}

\begin{figure}
\centering
\begin{tabular}{cc}
\includegraphics[width=0.4\textwidth]{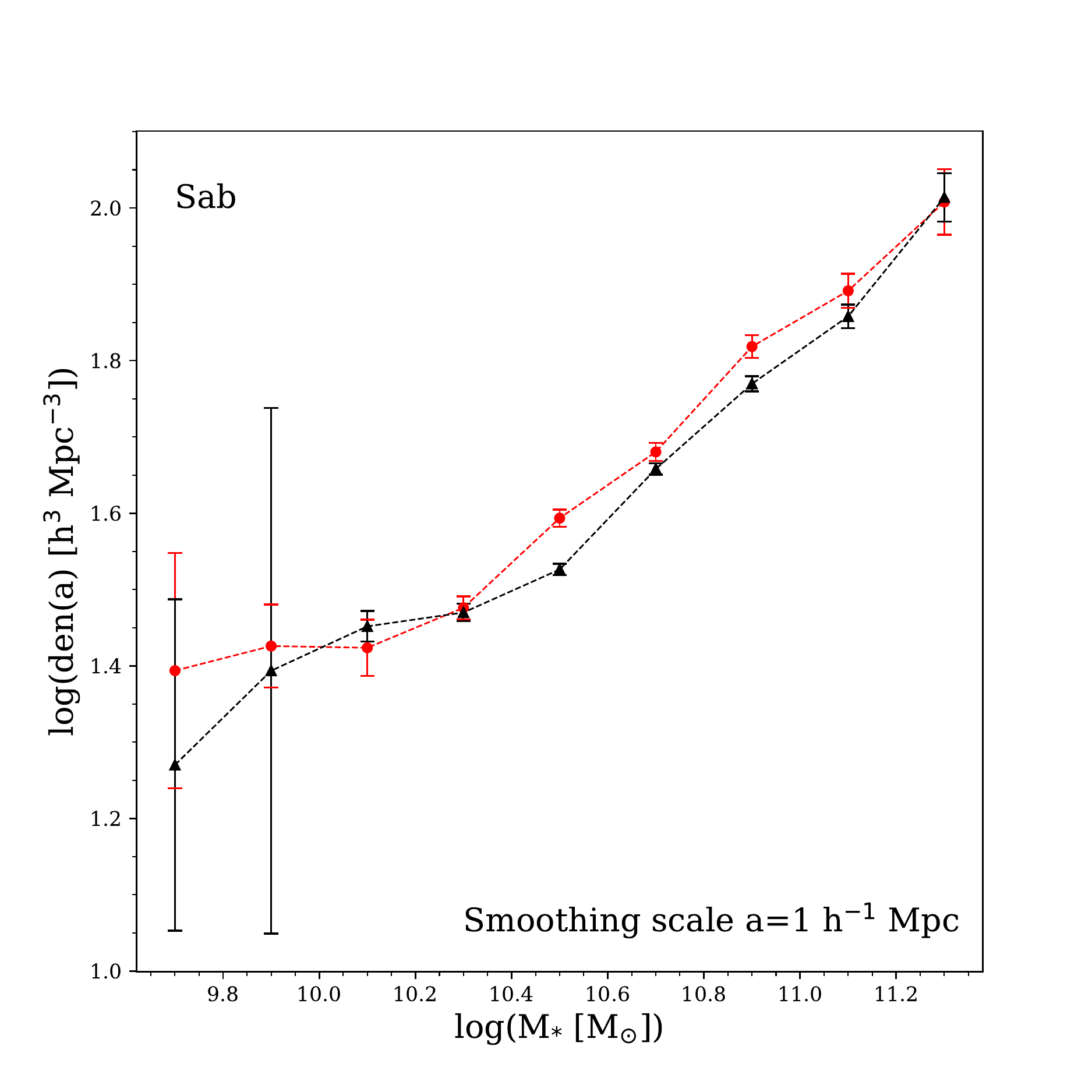}  \\ 
 \includegraphics[width=0.4\textwidth]{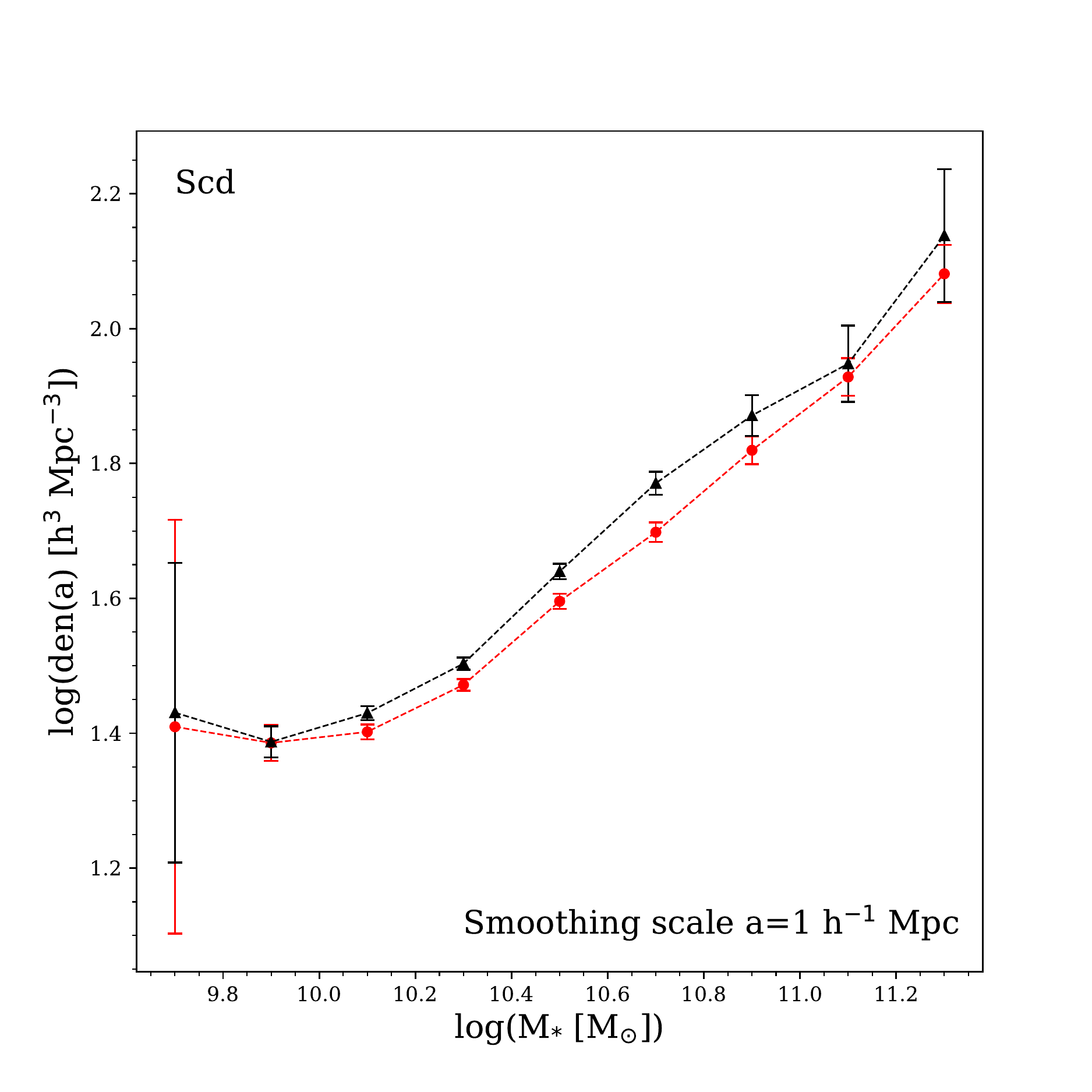}
\end{tabular}
\caption{Median value of the  local density as accounted by den($a$) as a function of stellar mass, segregating the VL sample between early- (top) and late-types (bottom). As for the previous figure, a smoothing scale of $a=1$ $h^{-1}$ Mpc is used.} 
\label{fig:morph-dena} 
\end{figure}

An other way to characterize the environment is looking at the local density. Table \ref{table:Tempel}, columns 2-4, shows the median value of the density at different smoothing scales (den($a$)) for the two different sub-samples. We note that at all scales, the environmental density is lower for LSBs than for HSBs. KS-tests suggest that these distributions are statistically different. In principle, these differences could be driven by differences in the subsamples, such as different stellar mass distributions or different morphological composition. In order to explore if  LSBs are found preferentially in low density environments, regardless of their stellar mass and morphology, we look at the local density as a function of stellar mass in Figure \ref{fig:dena}, and then splitting the sample into early- and late-types in Figure \ref{fig:morph-dena} (we focus on the density for a smoothing scale of $a=1$ $h^{-1}$ Mpc, with similar results found with the other smoothing scales). When controlling only for the stellar mass, we do not find significant differences, but once we distinguish between morphological types, for the case of late-type galaxies, LSB ones tend to reside in lower density environments compared with their HSBs counterparts, a result that is in line with previous findings based on number counts \citep{Rosenbaum09} and clustering \citep{DavisDjor85, Mo94}. 

We compare these results with the overdensities taken from \citet{Jasche10}. In the last column of Table \ref{table:Tempel} we show the median value of ln(2+$\delta$). Again, we observe marginal but significant differences between LSB and HSB galaxies, showing that LSB galaxies trend to be found in lower density environments. We therefore conclude that the large-scale structure has lower impact than the local environment, on determining if a galaxy becomes a LSB or an HSB one.

 \subsection{Dark matter Halo mass}
 
\begin{figure}
\centering
\includegraphics[width=0.4\textwidth]{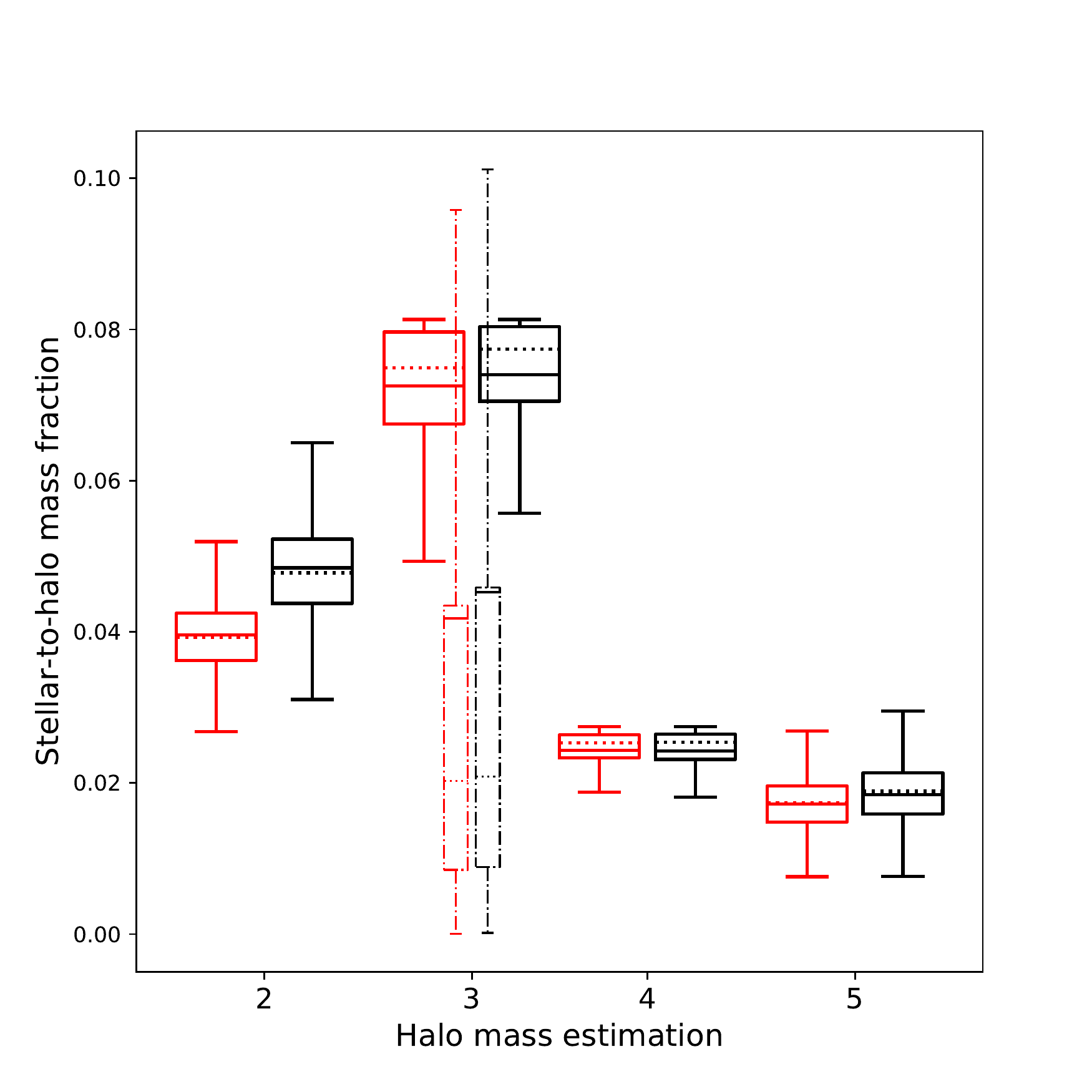}
\caption{Mean (solid lines) and median (dotted lines) for the stellar-to-halo mass ratio distributions of LSB and HSB galaxies of the VL sample, corresponding to the $f_{2,3,4,5}$ dark matter halo mass estimators. Dashed-dotted lines correspond to the $f_3$ distributions of the CS. Error bars indicate 1.5 times the inter-quartile range (IQR) of the corresponding distribution.}
\label{fig:fn} %CHECK!!!
\end{figure}

From the different halo mass estimates described in sec. \ref{sec:DMhalo}, we are able to look at the stellar-to-halo mass fractions for the galaxies in our VL sample, and compare the distributions of these ratios ($f_n$ with $n=$ 2, 3, 4 and 5 according to the halo mass estimate) for the HSB and LSB sub-samples. It is important to point out that we will only compare stellar-to-halo mass fractions between LSB and HSB galaxies within the same halo mass estimate, as the estimates rely on very different assumptions and methods to assign masses to the haloes associated to each galaxy. Figure \ref{fig:fn} presents box plots enclosing 50\% of the data. The solid line-type boxes indicate $M_*/M_H$ for the VL sample, whereas dash-dotted line boxes correspond to the CS. In this case, error bars correspond to 1.5 times the inter-quartile range (IQR) of the distribution. Inside each box-plot, solid and dashed lines correspond to the mean and median values of each distribution respectively. KS tests are applied to compare HSB and LSB samples, obtaining in all cases $p-$values lower than 0.001, which allows us to reject our null-hypothesis that both sub samples come from the same parent distribution. In all cases the LSB sub sample presents systematically lower stellar-to-halo mass ratios than the HSB sample.

Of the halo mass estimators we are implementing, the only one that uses kinematic data of the galaxies is $f_3$, which relies on an abundance matching procedure between the dark matter halo and galaxy circular velocities. Given that we count with kinematic information form ALFALFA for our CS, we show in the same figure the $f_3$ the mean and median values of the distributions for this sample, with the LSB sample presenting again lower values the their HSB counterpart. We observe that these results are consistent independently of the method employed, for all the estimations, the value of $f_n$ for LSBs is lower than for HSBs, with a difference reaching up to $\sim22\%$, in general agreement with previous findings \citep{Pickering97,McGaugh01,Swaters03,KuzioSpekkens11}.

\subsection{Group galaxies and Assembly Time}

From the \citet{Yang07} galaxy catalogue, we identify if the galaxies in our VL sample belong to galaxy groups or if they are isolated, and if they belong to a group, the catalogue specifies if the galaxy is the central one or if its a satellite, by ranking them according to their stellar masses, the most massive one of each group being the central one. Table \ref{table:groups} presents the membership to galaxy groups for our two VL sub-samples, showing a slightly higher fraction of isolated galaxies for the case of LSB over HSB galaxies, and other evidence showing that LSB galaxies are preferentially formed in relatively isolated environments.

\begin{figure}
\centering
\begin{tabular}{c}
\includegraphics[width=0.45\textwidth]{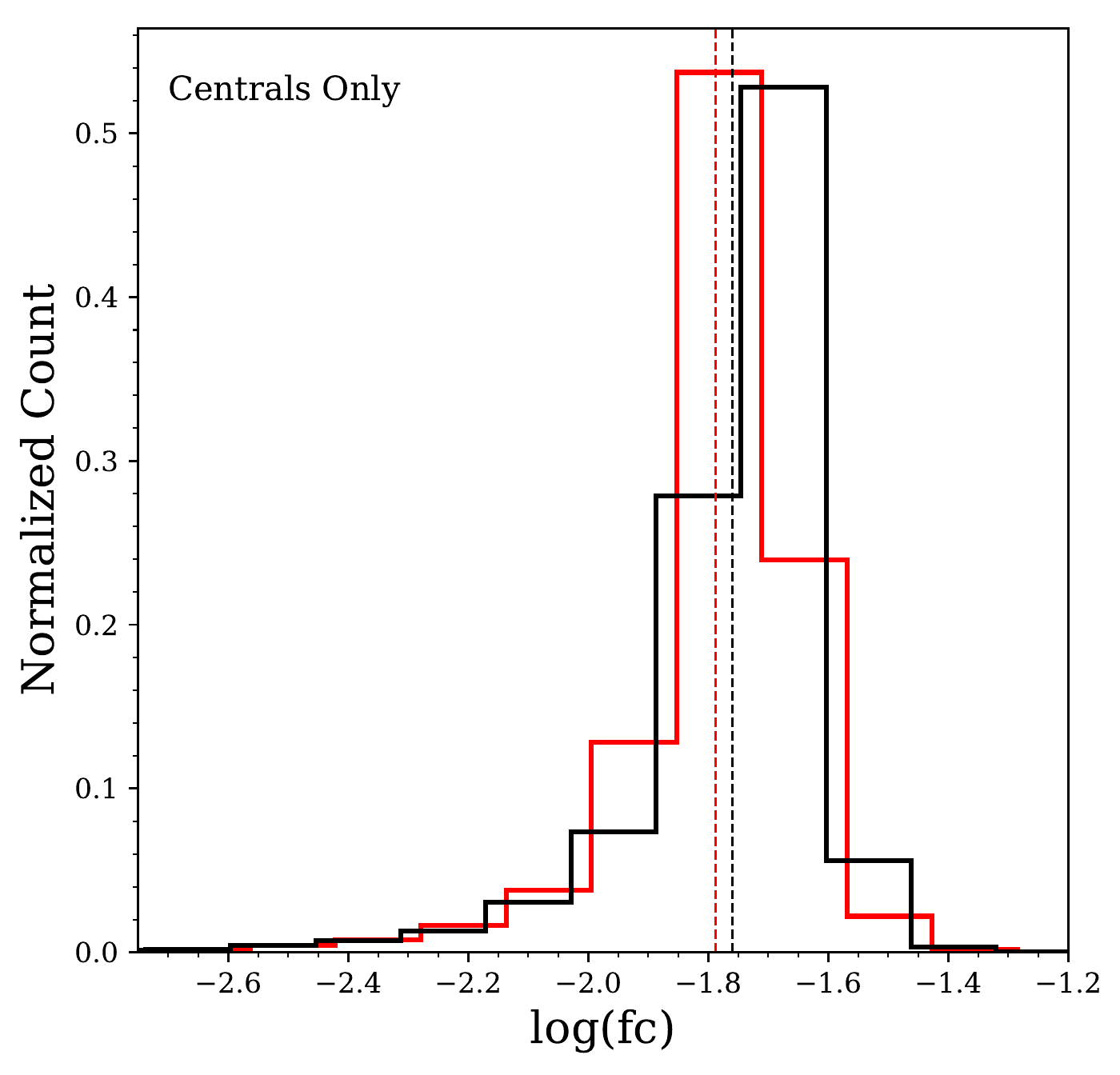} \\
 \includegraphics[width=0.47\textwidth]{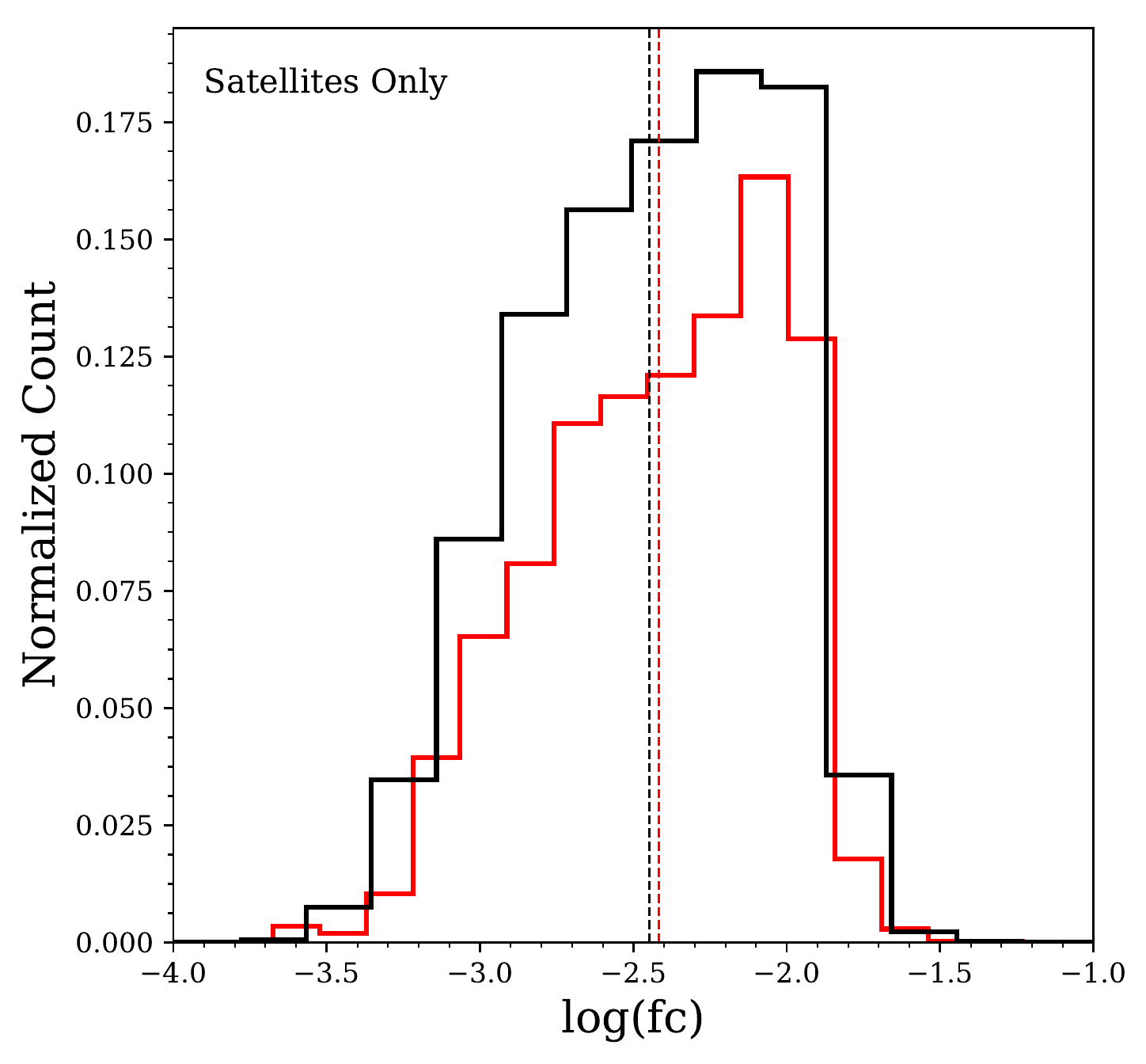}
\end{tabular}
\caption{\textit{Top:} Distributions of the proxy of assembly time for central LSB and HSB galaxies (red and black lines, respectively) of the volume limited sample.\textit{Bottom:} Same figure, but for satellite galaxies. In both cases, vertical dashed lines represent the mean value of log($f_c$) with the corresponding colour convention}
\label{fig:logfc}
\end{figure}

\begin{figure}
\centering
\includegraphics[width=0.5\textwidth]{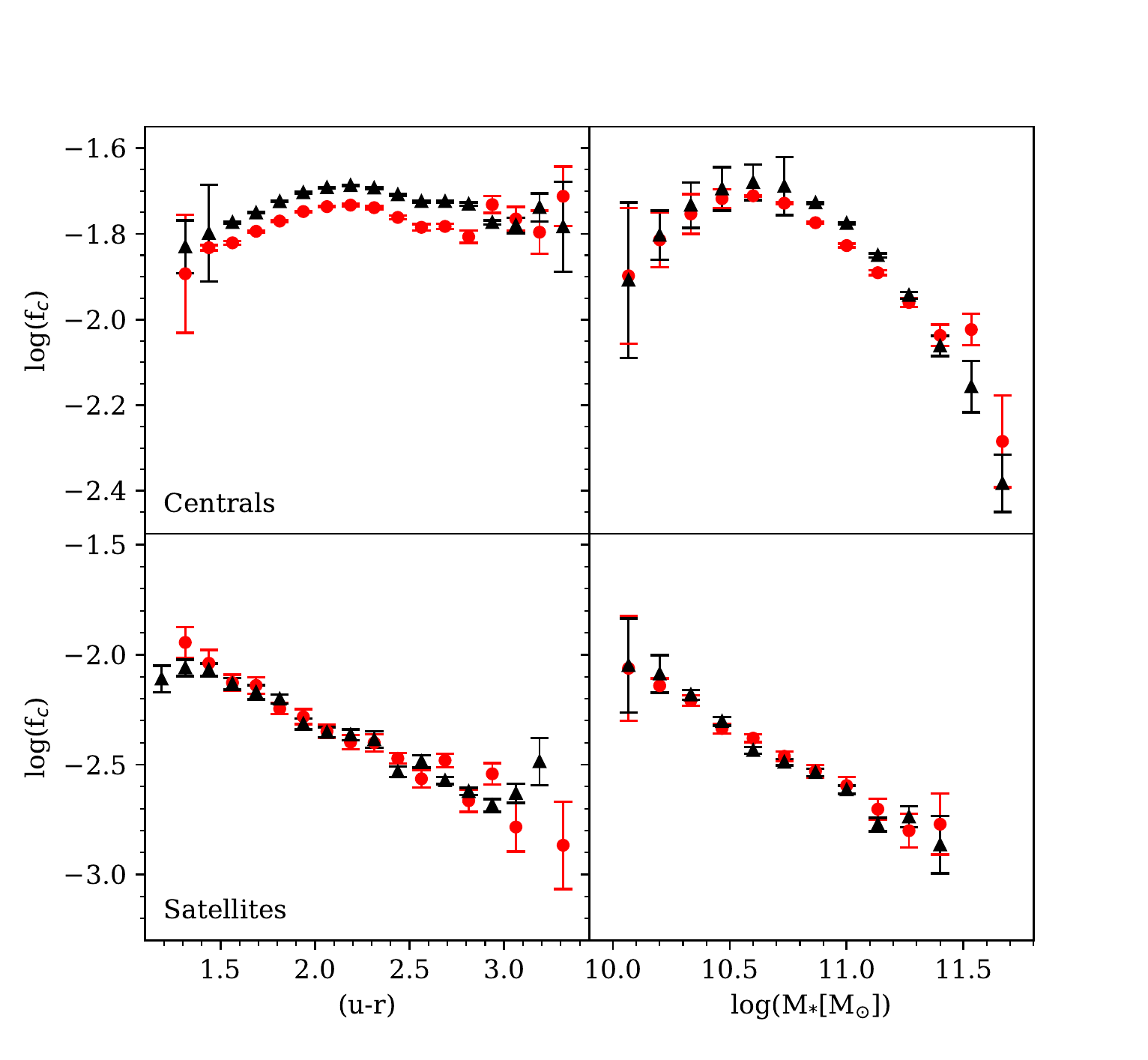}
\caption{Median value of the halo assembly time proxy $f_c$ as a function of colour (left) and stellar mass (right), for central (top) and satellite (bottom) galaxies. }
\label{fig:CentSat} %CHECK!!!
\end{figure}

\begin{table}
\centering % used for centering table
\begin{tabular}{c c c} % centered columns
\hline\hline 
Type & LSB &HSB \\ [0.5ex] % inserts table
%heading
\hline % inserts single horizontal line
Isolated & 67.9\% & 62.3\%\\ 
Central  & 10.1\% & 13.7\%\\ 
Satellite  & 22\% & 24\% \\ [0.5ex] 
\hline
\end{tabular}
\caption{Percentage of membership to galaxy groups.}
\label{table:groups} %CHECK!!!
\end{table}

Employing the same group catalogue, we are able to explore the evolutionary state of the galaxies in the VL sample by using $f_c$ as a proxy for the halo assembly time, with higher values of $f_c$ indicating earlier assembly times. Figure \ref{fig:logfc} presents $f_c$ distributions for central and satellites galaxies in the sample, showing that central LSB galaxies presents lower $f_c$ values than central HSB ones, while for satellite galaxies the opposite is found, although the difference is smaller. 

In their study,  \cite{Lim16} found that $f_c$ correlates with galaxy properties such as colour, stellar mass and star formation. Given that the colour and stellar mass distributions of LSB and HSB galaxies in the VL sample are different, the differences we detect in $f_c$ in Figure \ref{fig:logfc} could be indirectly driven by these other properties instead of the surface brightness. In Figure \ref{fig:CentSat} we test if at fixed stellar mass or colour, the value of $f_c$ is different for our two sub-samples. For the case of central galaxies, at fixed colour, LSBs present lower values of $f_c$, which indicates later assembly times than for the case of HSB galaxies. At fixed stellar mass the difference between low and high surface brightness galaxies is smaller, but for log(M$_*$) $<$11.25, LSB galaxies present systematically lower values than HSB galaxies. These results indicate that for the case of central galaxies, LSBs assembled half of their total mass at later times than HSBs, supporting the idea that LSB galaxies are less evolved systems than their HSB counterparts. This result is of particular interest as it is well known that LSB galaxies have younger stellar populations than HSB ones, but this result indicates that not only the star formation rate occurs at lower rate, but also the halo is also assembled later.

For the case of satellite galaxies (Figure \ref{fig:CentSat}, lower panels), no noticeable, systematical difference is found for the value of $f_c$ between LSB and HSB galaxies, not at fixed colour, nor at fixed stellar mass.

\subsection{Specific Angular Momentum and Spin}

\begin{figure*}
\centering
\begin{tabular}{cc}
\includegraphics[width=0.4\textwidth]{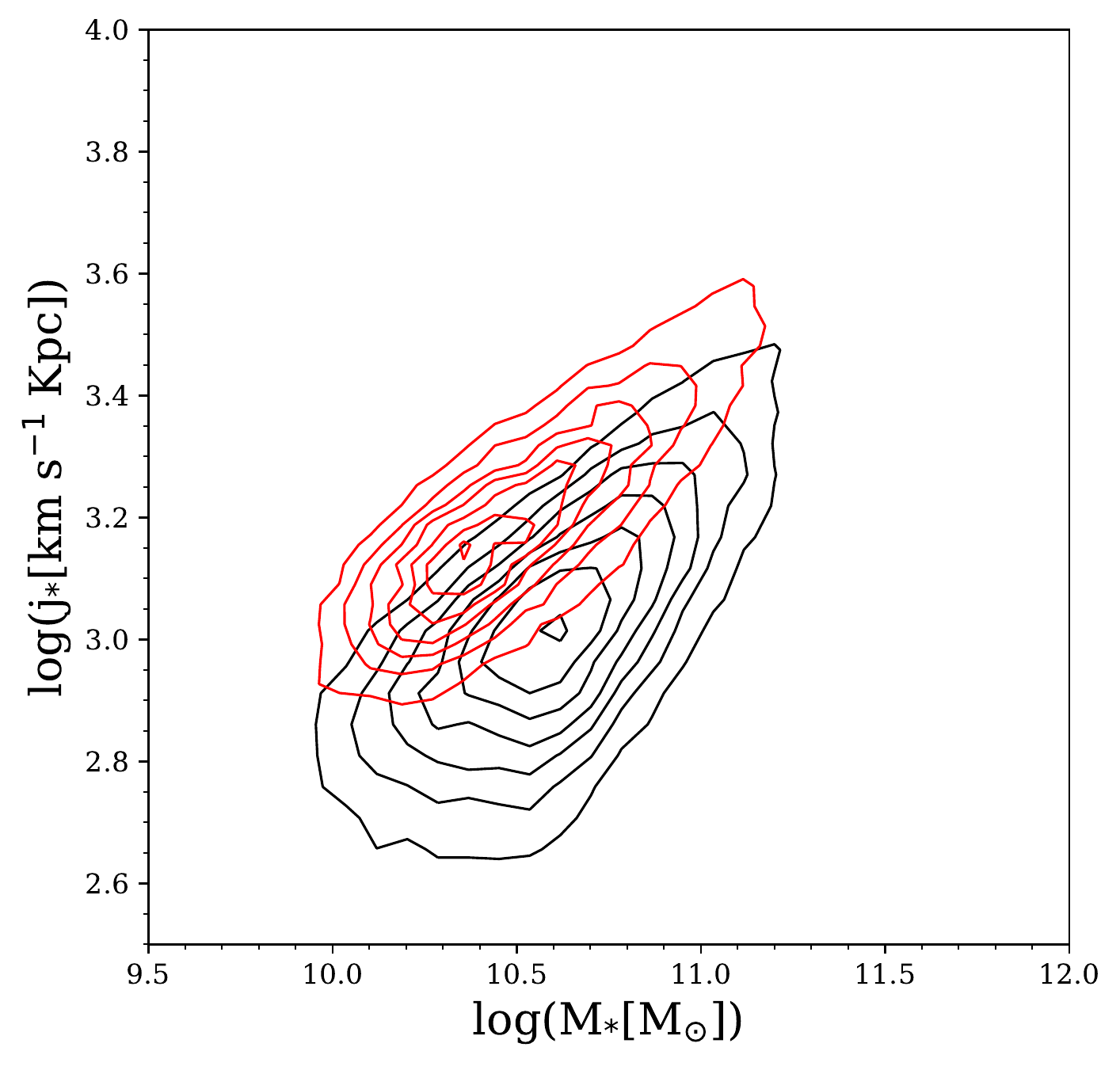} & \includegraphics[width=0.4\textwidth]{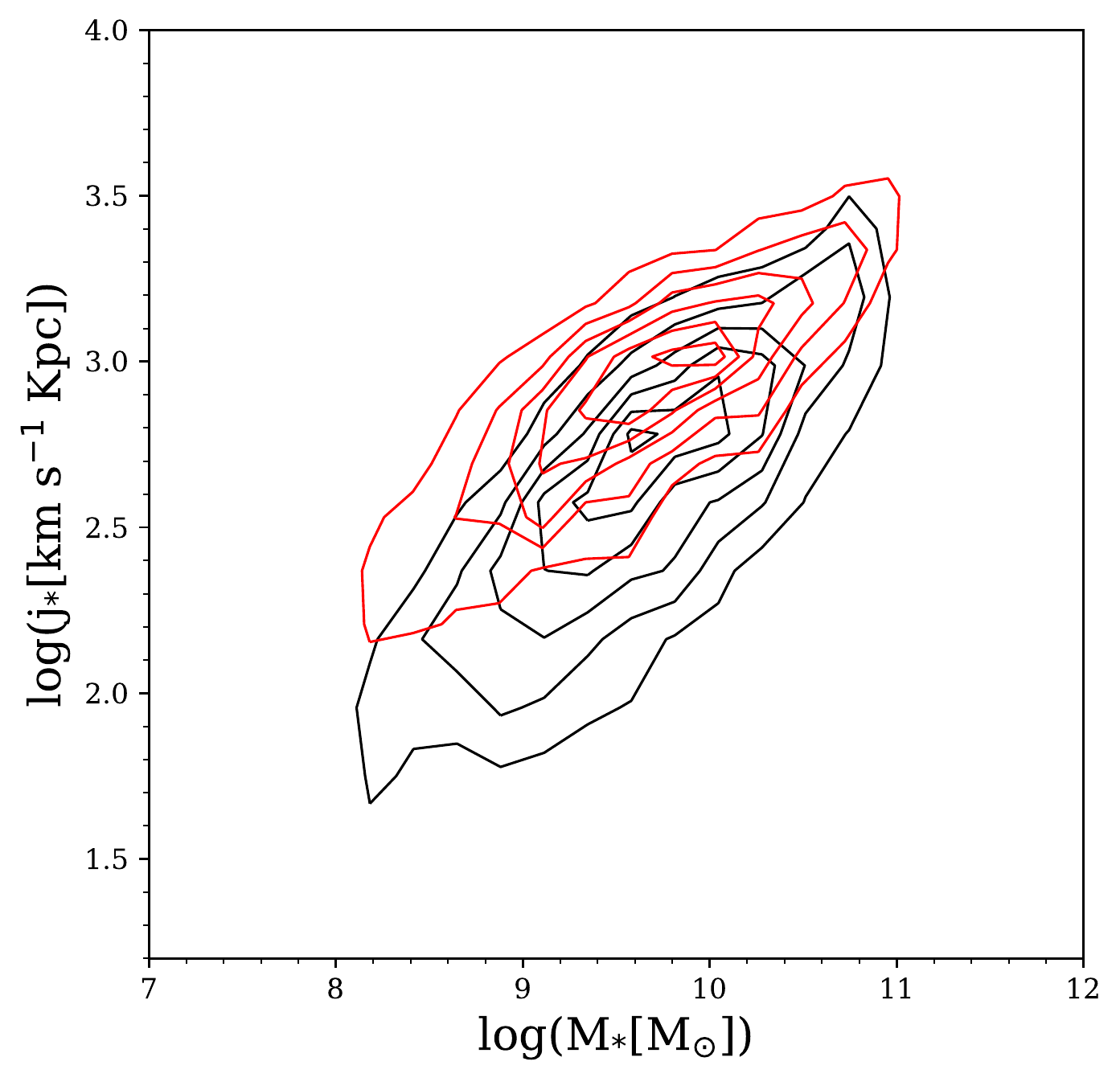}\\
\includegraphics[width=0.4\textwidth]{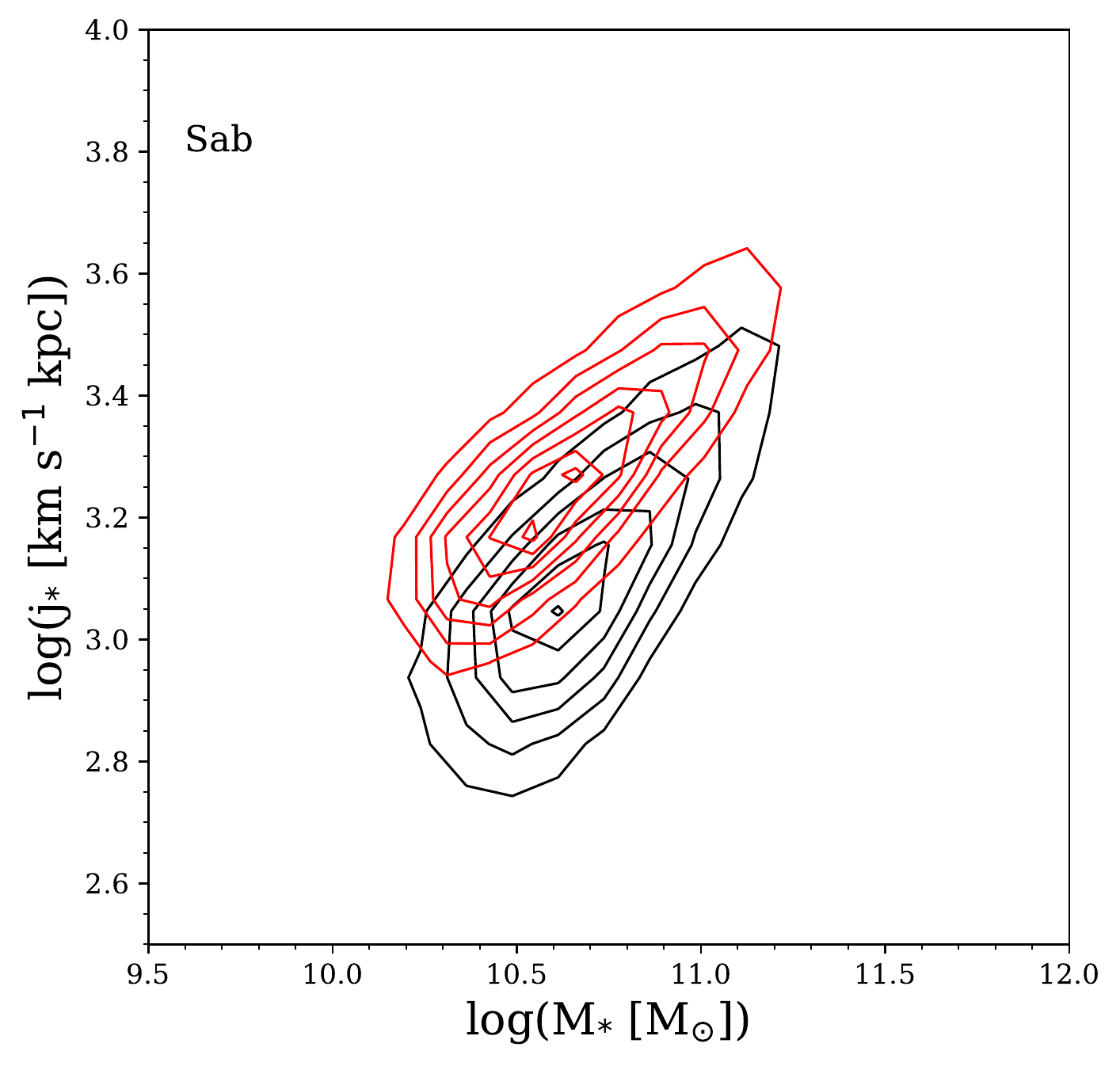} & \includegraphics[width=0.4\textwidth]{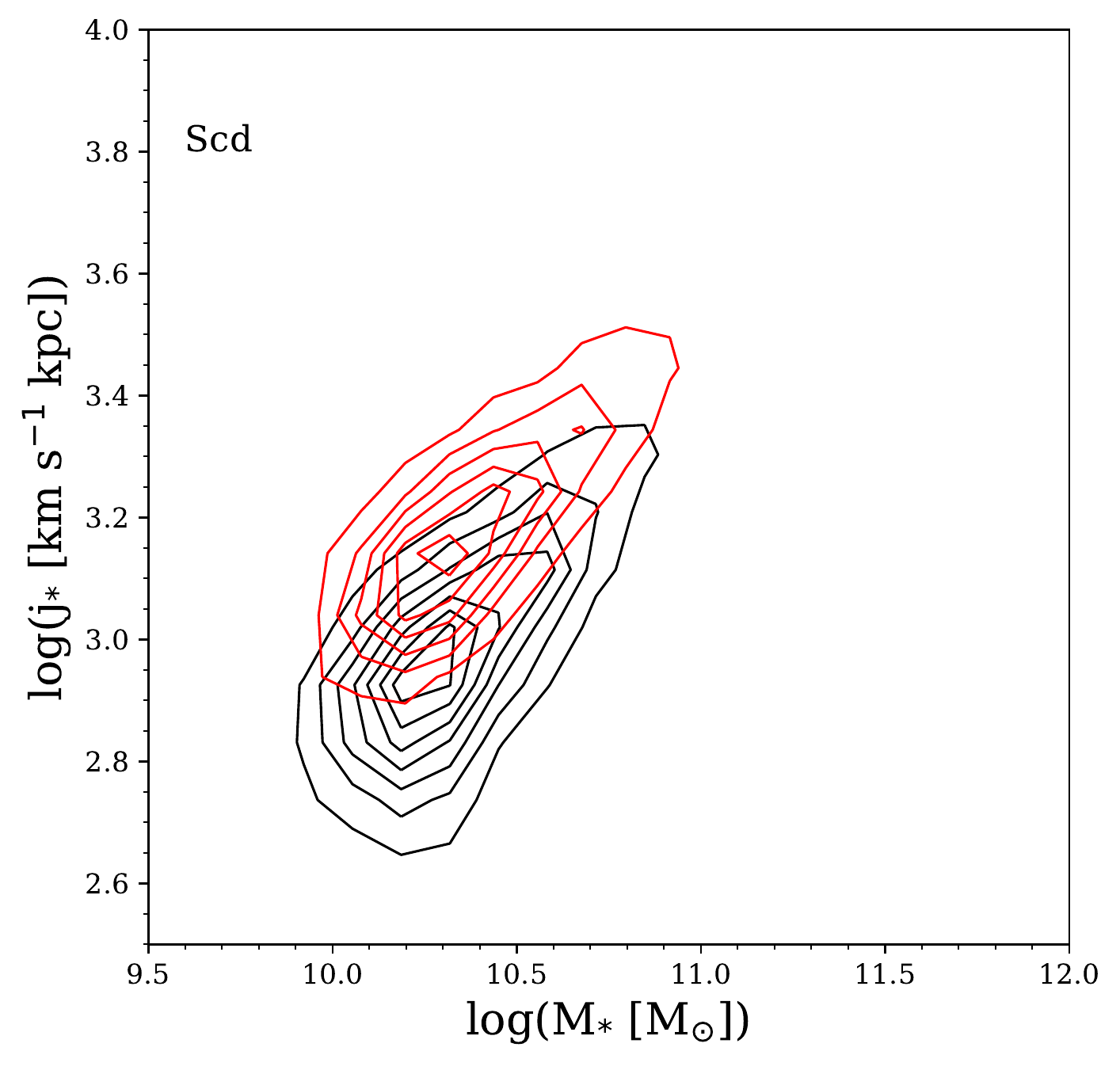}
\end{tabular}
\caption{\textit{Top panels}: Density contours of the specific angular momentum j$_{*}$ as a function of stellar mass M$_*$ for LSB (red) and HSB (black) galaxies in the VL (left panel) and CS (right panel) samples. \textit{Bottom panels}: Density contours of the specific angular momentum j$_{*}$ as a function of stellar mass M$_*$ for the VL sample, segregated into early (left panel) and late-types (right panel).}
\label{js} %CHECK!!!
\end{figure*}

In order to investigate if LSB galaxies present larger values of the spin parameter than their high surface brightness counterparts, we start by exploring the differences in the specific angular momentum between the two types, as all of our spin estimators enlisted in Section \ref{sec:SpinCalculation} rely strongly on $j_{*}$. In Figure \ref{js}, top panels, we present the distribution of galaxies in the $j_{*}$ versus M$_{*}$ diagram, for  VL (left) and CS (right), with each density contour enclosing $\sim$15\% of the data. It is clear that, for the VL sample, the distribution of LSB galaxies in this plane is shifted to higher values of $j_*$ and lower values of M$_*$, when compared with HSB galaxies, as expected. For the case of the CS, given that both sub-samples present the same stellar mass distribution, the shift is only present in the vertical direction. With the CS we are able to explore a larger range in mass, as we are not limited to intrinsically bright and massive systems, but qualitatively the trend shows the same for both, VL and CS; at fixed stellar mass LSB galaxies present higher angular momentum as recently reported by \cite{Jadhav19} for a sample of superthin bulgeless LSB galaxies.

The early work by \cite{Fall83}, and later later revisions \citep{RomFall12}, have shown that ellipticals and late-type spirals follow parallel sequences in the $j_*$-M$_*$ plane, with ellipticals having a factor of $\sim$3-4 times lower $j_*$ at fixed $M_*$, with lenticular galaxies on average lying in between. As LSB galaxies tend to present late-type morphologies, the difference between in the distribution of LSB and HSB in the $j_*$-M$_*$ plane could be in part driven by morphology. To eliminate this dependency, we segregate our VL sample into early- and late-types in the bottom panels of \ref{js}, where we see that even controlling by morphological type, high and low surface brightness galaxies occupy different regions in this plane, with LSB presenting higher values of $j_*$ at fixed $M_*$.

\begin{figure*}
\centering
\begin{tabular}{cc}
\includegraphics[width=0.4\textwidth]{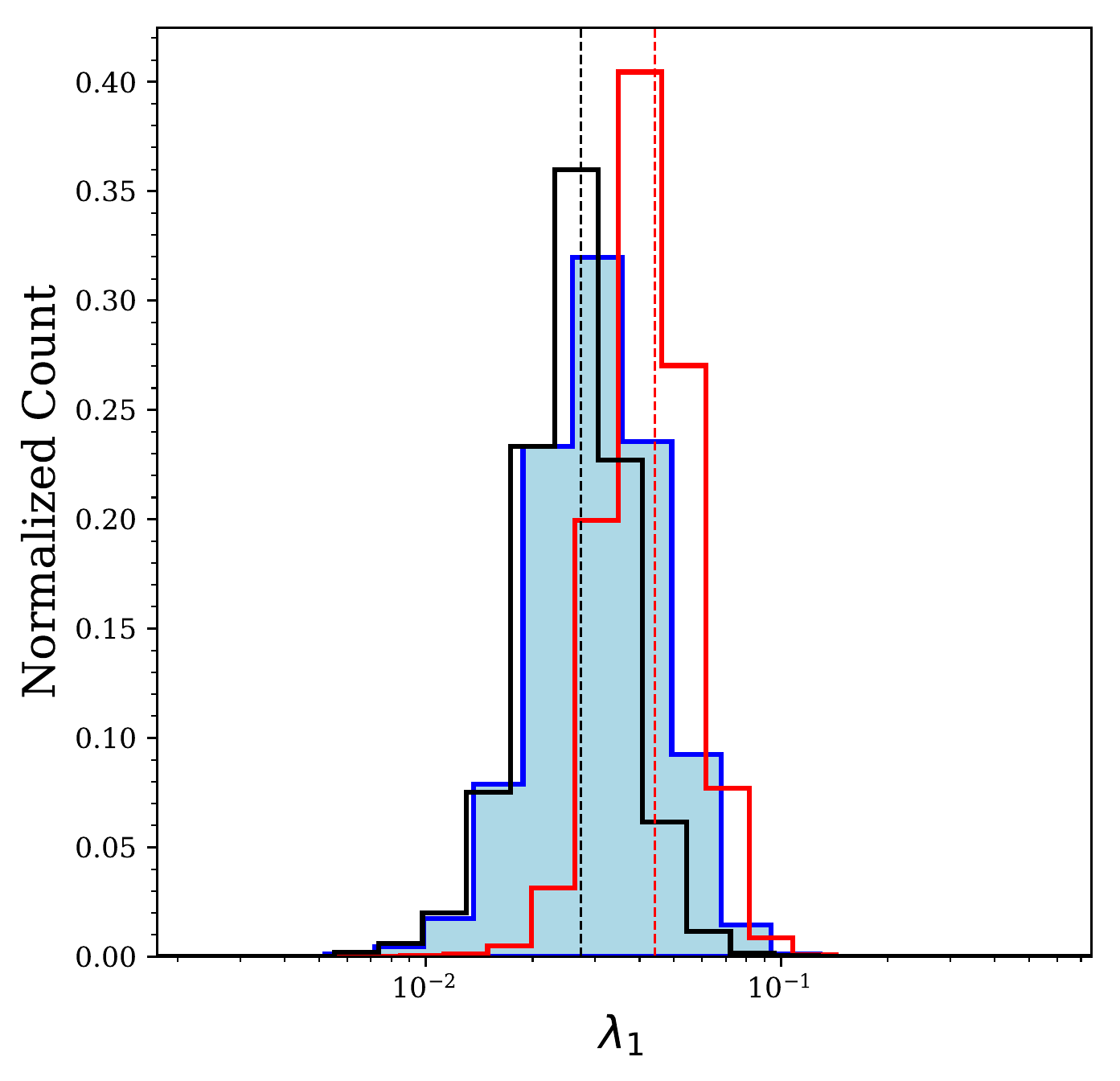} & \includegraphics[width=0.4\textwidth]{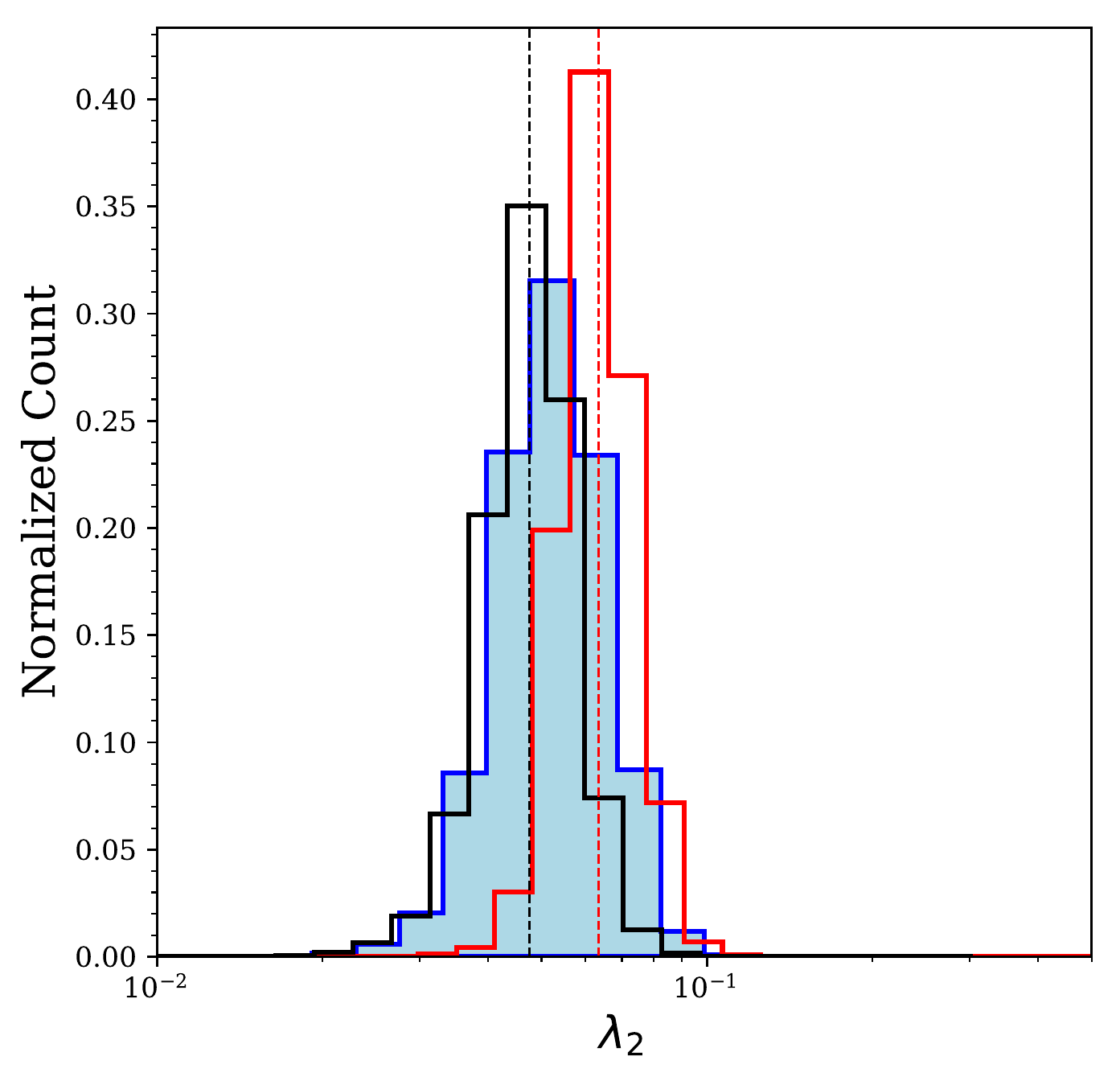} \\
\includegraphics[width=0.4\textwidth]{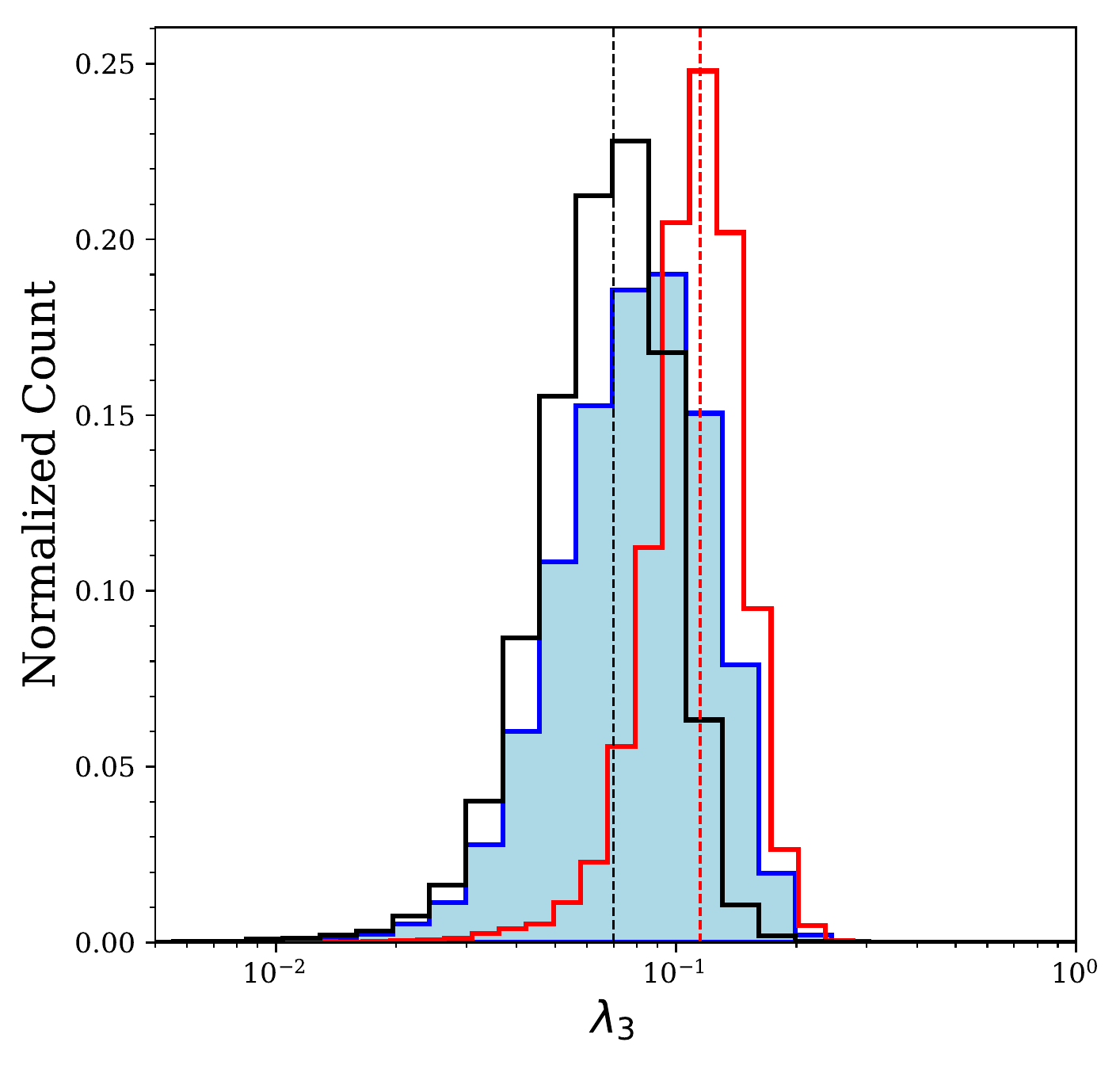} & \includegraphics[width=0.4\textwidth]{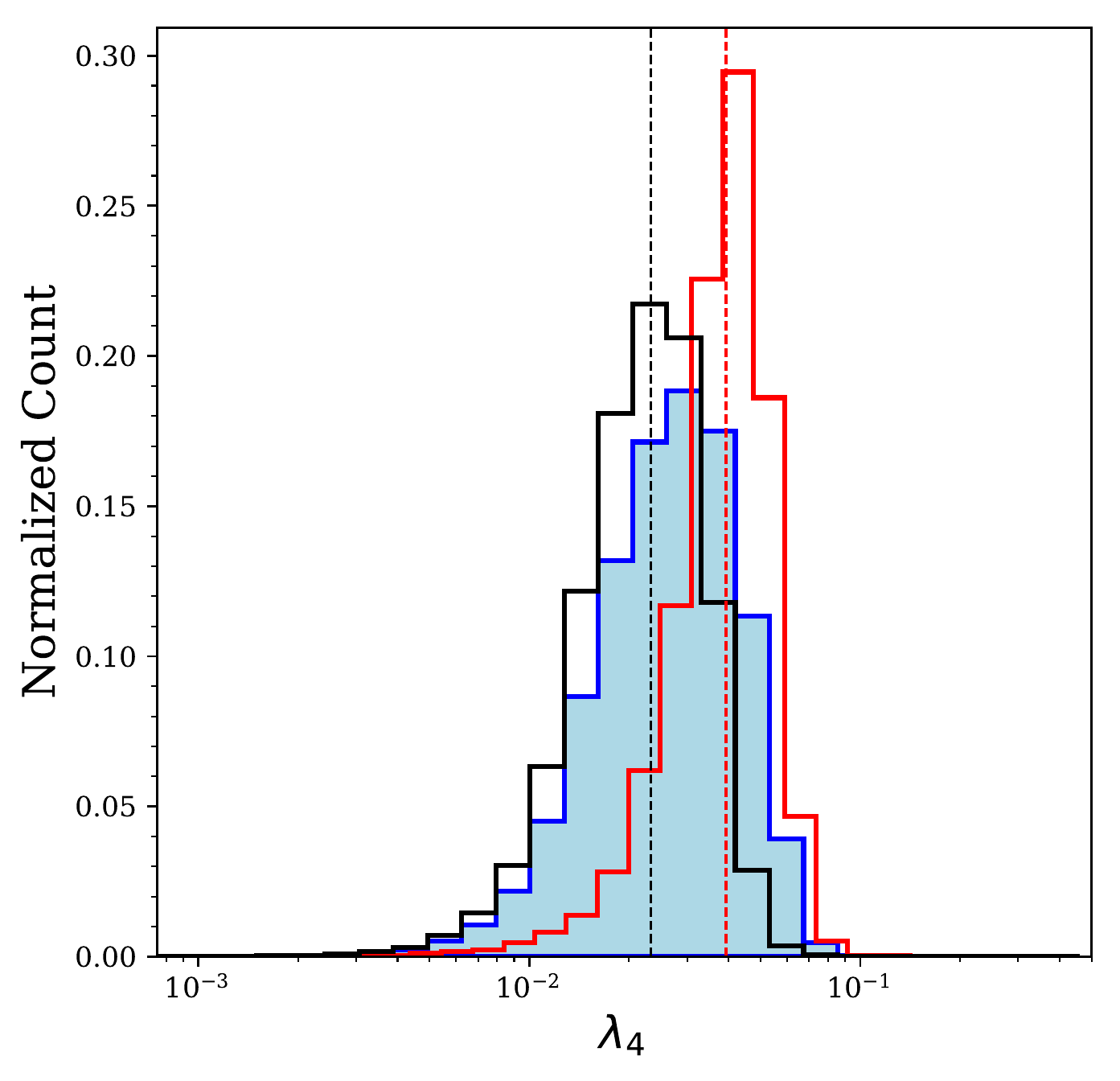} \\
\includegraphics[width=0.4\textwidth]{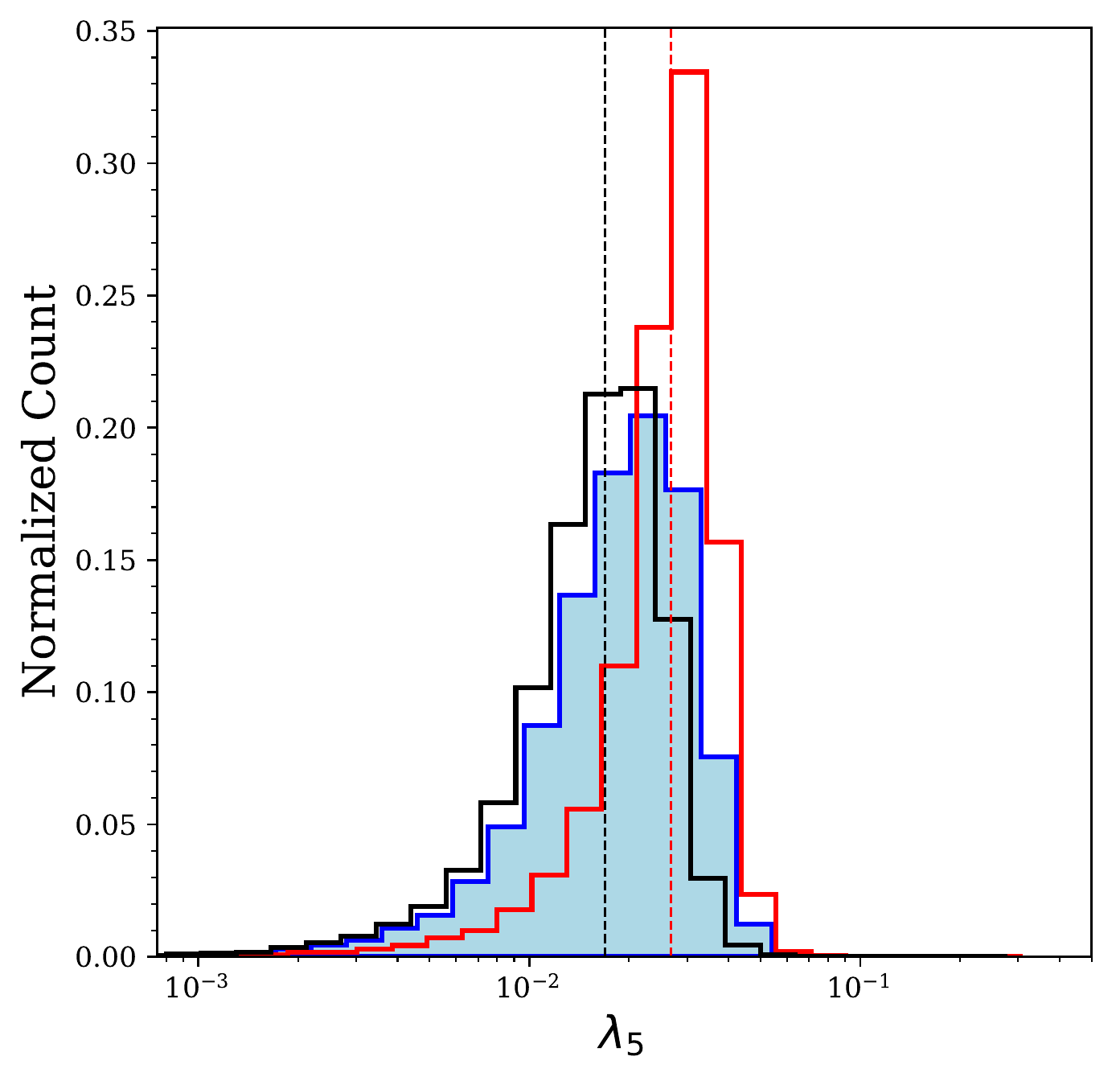} & \includegraphics[width=0.4\textwidth]{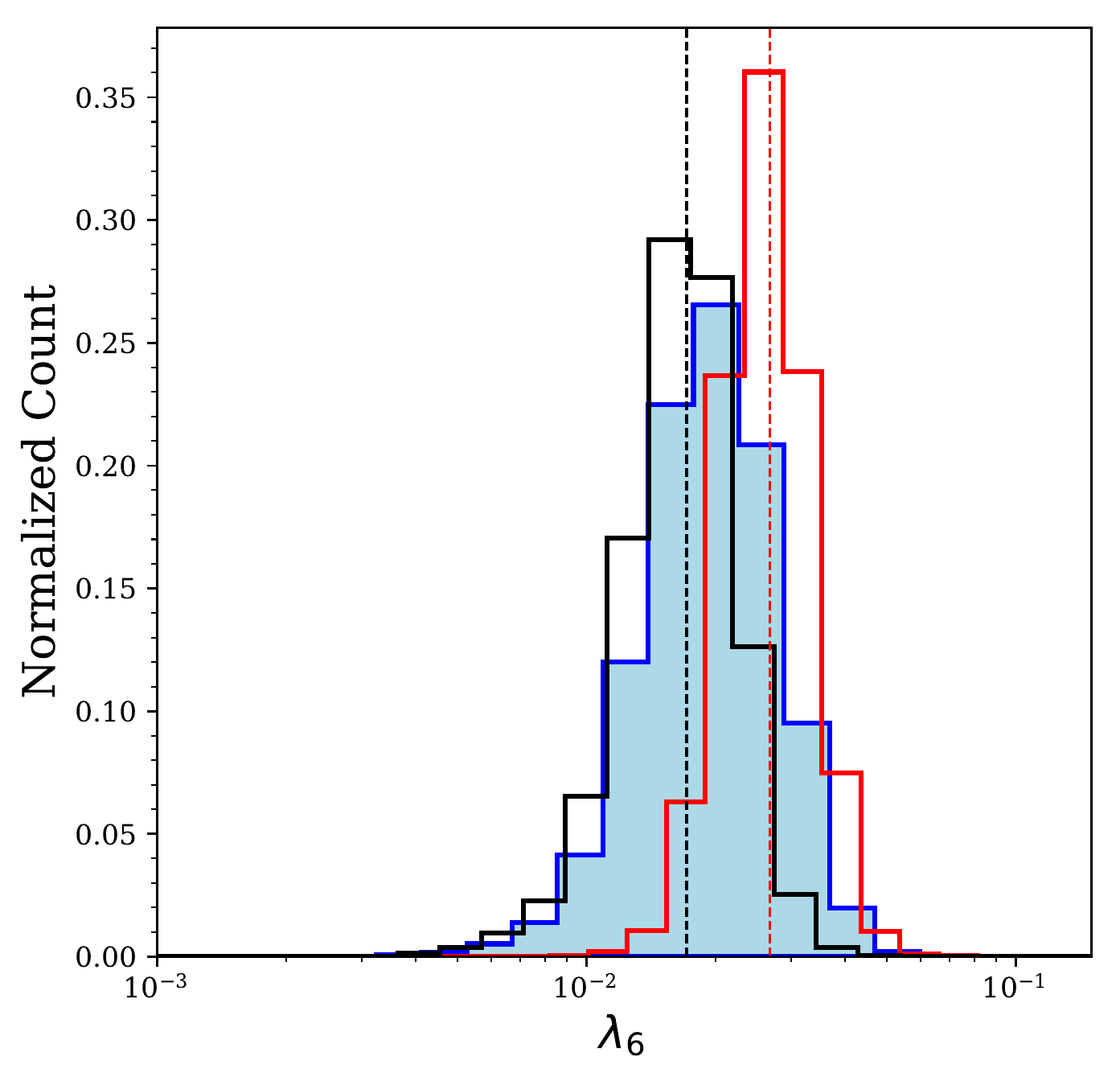} \\
\end{tabular}
\caption{Spin distributions for the VL sample. The dashed line corresponds to the mean value of $\lambda$ for LSB and HSB galaxies. The blue-shaded region corresponds to the full distribution of $\lambda$ using the five different estimations of M$_H$, plus the alternative expression adopted from \citet{Meurer18}.}
\label{VolLimSpins}
\end{figure*}

Finally, we are able to evaluate our six different spin estimators described in Section \ref{sec:SpinCalculation}. We start by comparing the spin distributions of LSB and HSB galaxies in our VL sample, where $V_{rot}$ is assigned through a TF relation. The distributions are displayed in Figure \ref{VolLimSpins}. Within a spin estimator, the hypothesis of the two sub-samples to be drawn form the same underlying distribution is discarded with a KS test. In all cases, the spin distribution of LSB galaxies is shifted to larger values when compared with the HSB sub-sample, with the mean value of the spin parameter of LSB galaxies being 1.33 to 1.70 times larger than the mean value obtained for HSBs.

\begin{figure*}
\centering
\begin{tabular}{cc}
\includegraphics[width=0.4\textwidth]{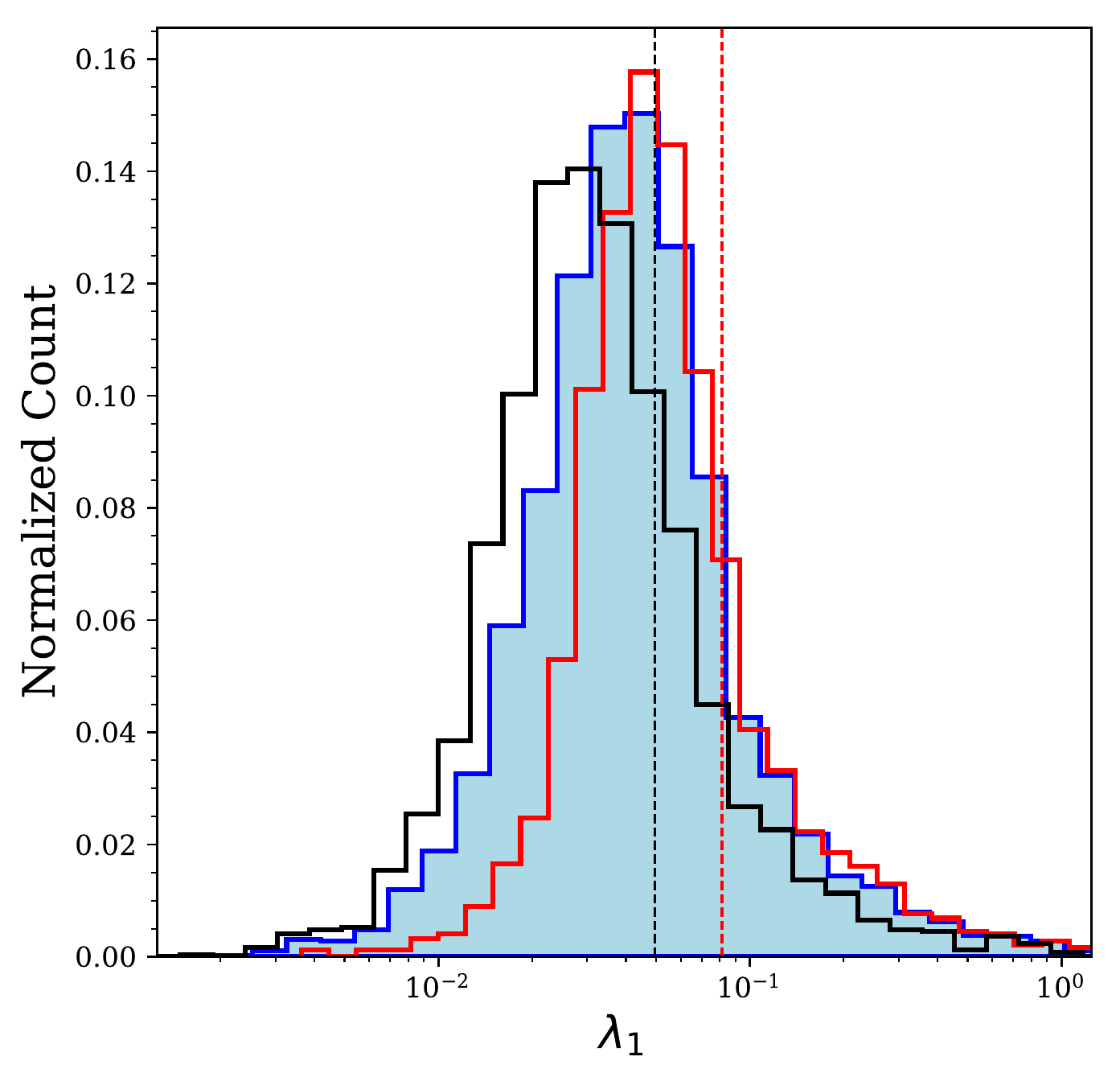} & \includegraphics[width=0.4\textwidth]{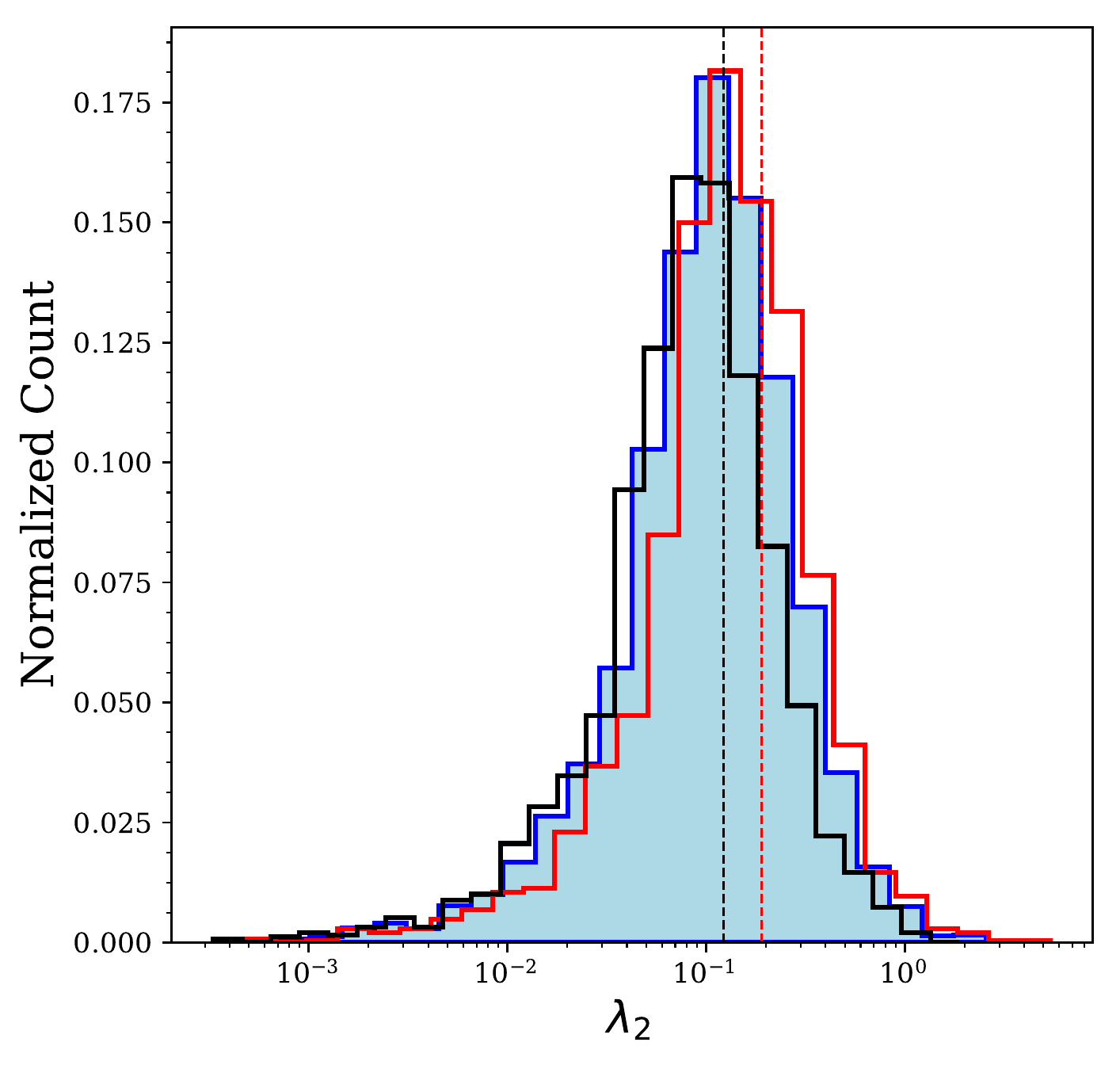} \\
\includegraphics[width=0.4\textwidth]{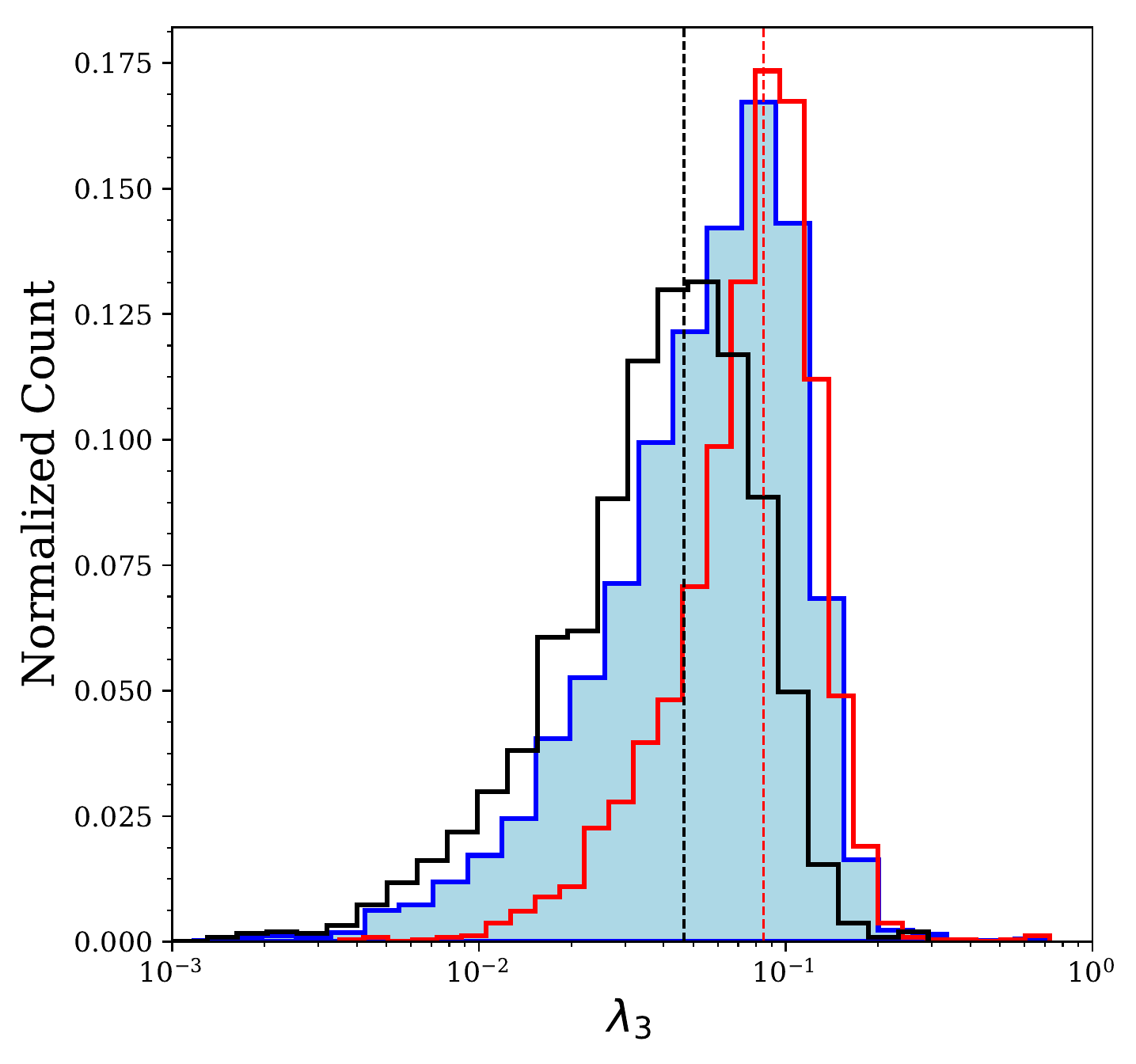} & \includegraphics[width=0.4\textwidth]{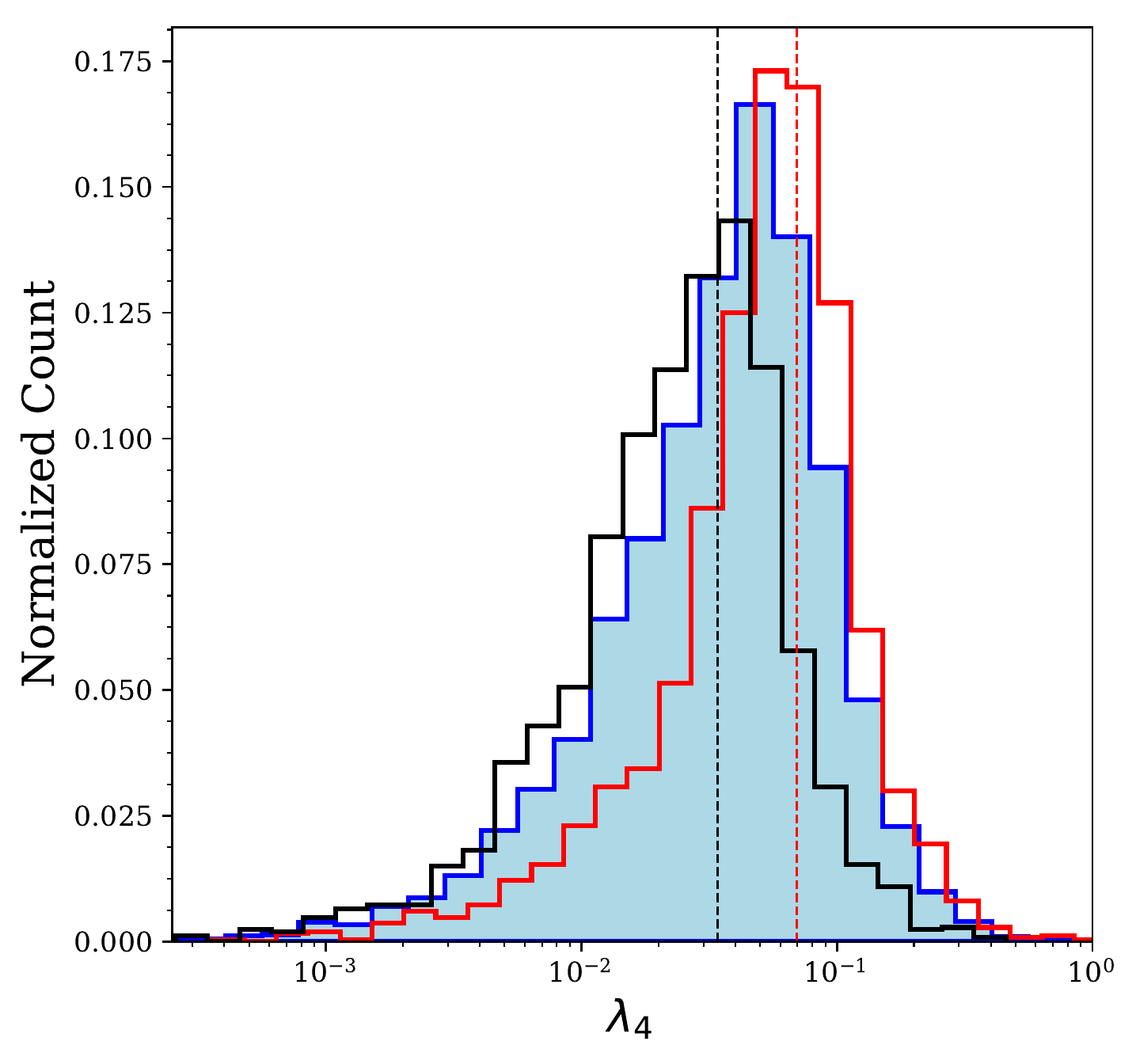} \\
\includegraphics[width=0.4\textwidth]{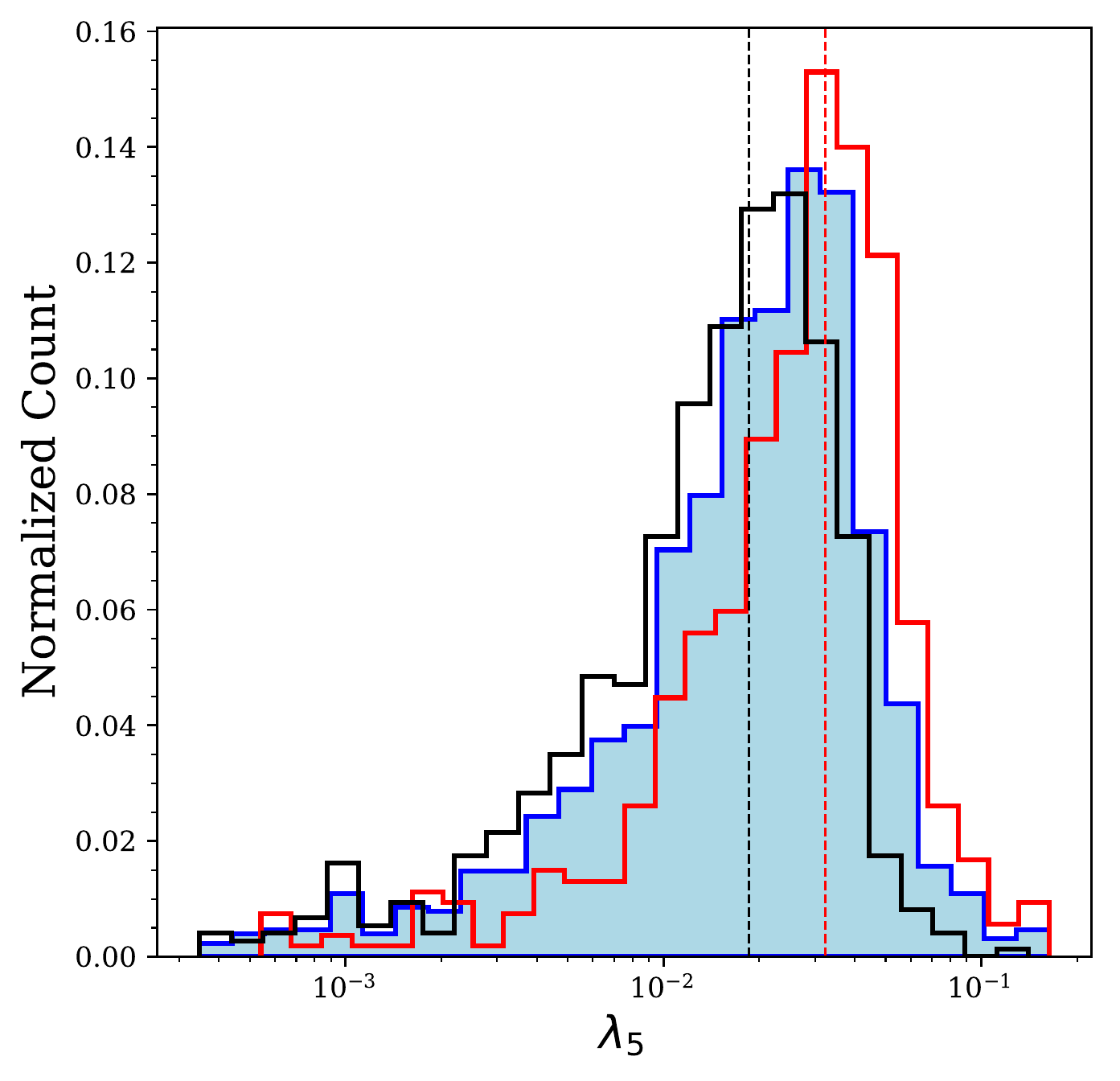} & \includegraphics[width=0.4\textwidth]{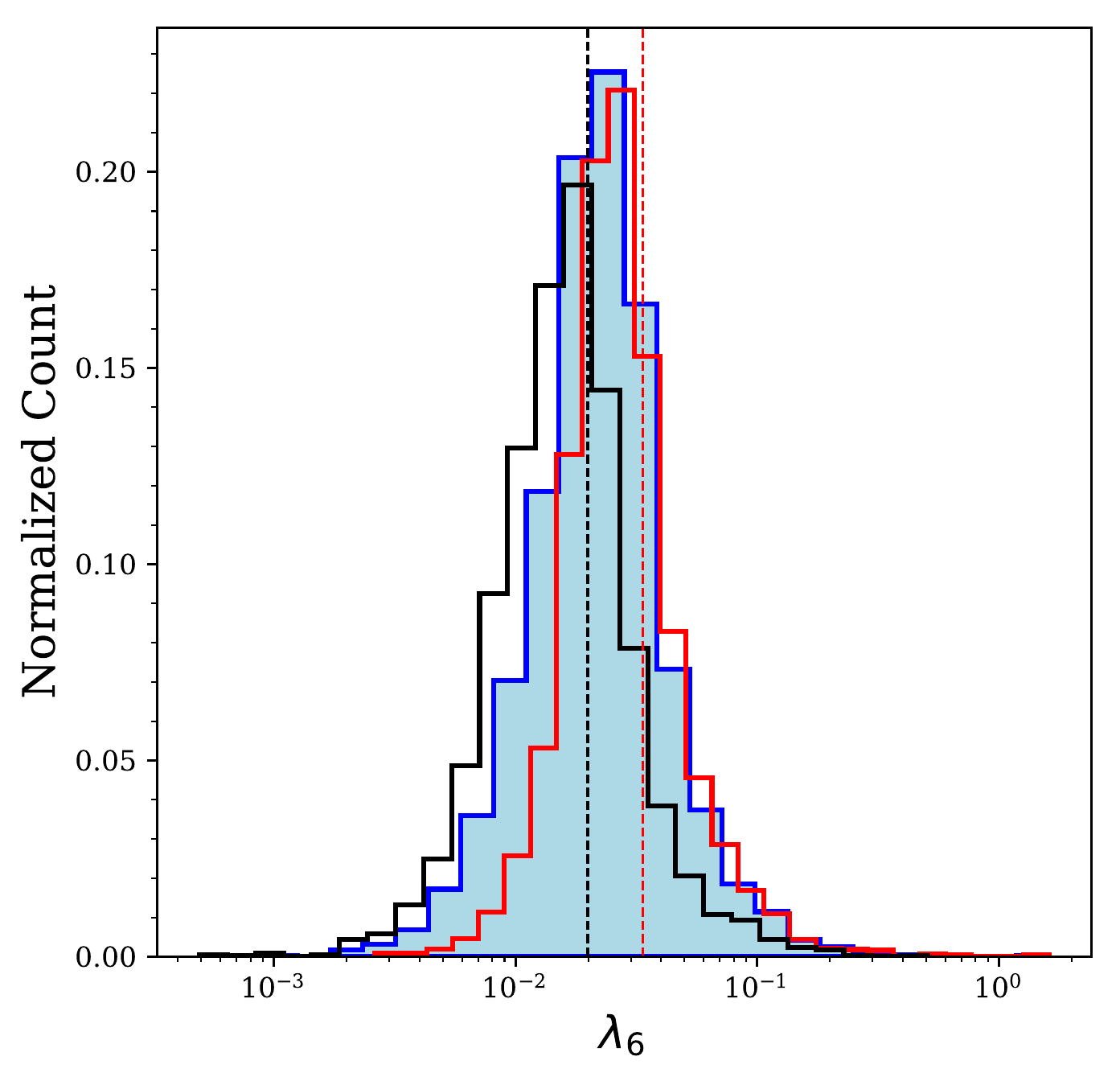} \\
\end{tabular}
\caption{Spin distributions for CS. The dashed line corresponds to the mean value of $\lambda$ for LSB and HSB galaxies. The blue-shaded region corresponds to the full distribution of $\lambda$ using the five different estimations of M$_H$.}
\label{ContSampSpins}
\end{figure*}

Although LSB galaxies follow the same Tully-Fisher relation as normal spiral galaxies \citep{Zwaan95}, we also present the same distributions using the CS for which we have a direct measurement of $V_{rot}$ form the kinematic information of ALFALFA. Given that our CS is constructed requesting that the two sub-samples share the same underlying distribution of $M_*$, we are already discarding a difference in $\lambda$ imposed by a difference in $M_*$. The corresponding distributions are displayed in Figure \ref{ContSampSpins}, where we see again the $\lambda$-distributions of LSBs systematically shifted to higher values when compared with the $\lambda$-distributions of HSB galaxies. Using the CS, the mean value of the spin parameter of LSB galaxies is 1.62 to 2 times larger than the one found for HSB galaxies.

The results presented in Figures \ref{VolLimSpins} and \ref{ContSampSpins} support the common knowledge \citep{Dalcanton97,Boissier03,KimLee13,Peschken17} that LSB galaxies are formed in fast-spinning dark matter haloes, and give constrains of the relative value of this parameter between LSB and HSB galaxies, setting this parameter as the one responsible for establishing their low surface brightness nature. 

We remind the reader that for the $\lambda$ estimates implemented in this work we assumed conservation of angular momentum between the components of the system, and that $j_*=j_H$, which is not strictly what is reported in theoretical works using numerical simulations. High resolution hydrodynamic simulations find that the specific angular momentum of the baryonic component is considerably higher than that of the dark matter halo \citep{Stewart11,Kimm11,Stewart13,Zjupa17}, which would set our $\lambda$ estimates as upper limit values. On their behalf, \citet{Teklu15} and \citet{Zavala16} report a correlation between $j_{*}$ and $j_H$, with a strong dependence on the morphology of the galaxies, with disc dominated galaxies retaining most of their specific angular momentum, in line with the results by  \citet{Genel15} that find a angular momentum retention factor for disc-dominated galaxies close to $\sim$100\%, whereas in early-type (bulge dominated) galaxies is only 10-30\%. Given that our sample was chosen to include mostly disc-dominated galaxies, we would expect our hypothesis not to be unrealistic, and the difference in $\lambda$ between LSB and HSB galaxies to be a solid result.

This result supports the conclusions by  \citet{Cerv13} and  \citet{Cerv17}, who found a lower fraction of barred galaxies between LSB galaxies when compared with HSB, and concluded that this was the effect of LSB galaxies having high spinning haloes that suppress the growth, in strength and size, of the bar, as predicted by numerical simulations of isolated systems \citep{Long14,Fujii19,Valencia19}.

\section{Summary and Conclusions}
\label{sec:Conclusions}

In the present work we made use of a volume-limited sample of galaxies drawn from the SDSS-DR7 to study the environment at different scales of LSB galaxies and compare it with that of HSB galaxies. When kinematic data was required to characterize the local environment of the galaxies, we made use of control samples constructed using HI kinematic data from ALFALFA. Our main findings are:

(i) Regarding the membership of galaxies to large scale structures, we find a slight preference of LSB galaxies to be found in filaments instead of clusters when compared with HSBs, and when we look at the distance of the galaxies to their nearest filament, for the case of early-type galaxies at fixed stellar mass, LSB are found farther than their HSB counterparts. Analysing the local density we find that LSB are found in lower density environments, specially if we for the case of late-type galaxies at fixed stellar mass. 

(ii) Using five different stellar-to-halo mass estimates we find that this ratio is systematically lower for LSB galaxies, with a difference reaching up to $\sim22\%$, indicating that LSB galaxies are more dark matter dominated than HSB galaxies.

(iii) From the group catalogue we identify that the fraction of isolated central LSB galaxies is higher than the fraction of HSB galaxies, showing that they inhabit more isolated environments. Using the same catalogue we find that for the case of central galaxies, LSBs present lower values of $f_c$, which indicates a later assembly time than for the case of HSBs, reinforcing the idea that these kind of galaxies are less evolved systems.

(iv) The largest difference between the physical parameters explored in the present work is with the specific angular momentum and the spin. Using the VL sample, where V$_{rot}$ is assigned through a stellar TF relation, the distribution LSB galaxies in the $j_*$-M$_*$ plane is shifted to larger $j_*$ values when compared with the distribution of HSB galaxies, even when the sample is segregated according to morphological type. We confirm this result using the CS, where V$_{rot}$ is directly measured from the width of the HI line profile and, by construction, the LSB and HSB samples share the same stellar mass distribution.

(v) Implementing six different spin estimators, we obtained $\lambda$-distributions for LSB and HSB galaxies in the VL and CS, finding, in all cases, a larger value of $\lambda$ for LSB galaxies, with the mean value of the spin parameter of LSB galaxies being 1.6 to 2 times larger than the one found for HSB galaxies.

Our results indicate that the main differences in environment between LSB and HSB galaxies is on small scales, namely the low stellar-to-halo mass ratio and the high spin, which would indicate that these are the two main parameters responsible in establishing the low surface nature of these galaxies, along with a later halo assembly time.

\section*{Acknowledgements}

The authors thank the thorough reading of the original manuscript by the anonymous referee and his/her insightful comments that helped to improve the quality of the paper and clarify the results.
The authors also thank Sebastian S\'anchez, Xavier Hernandez and Rosa A. Gonz\'alez-L\'opezlira for helpful comments that helped to improve
the analysis of results.
The authors acknowledges financial support through PAPIIT project IA103517 and IA103520 from DGAPA-UNAM.
The work of L. E. P. M. is supported by a CONACyT scholarship.
Funding for the SDSS and SDSS-II has been provided by the Alfred P. Sloan Foundation,
    the Participating Institutions, the National Science Foundation, the U.S. Department of
    Energy, the National Aeronautics and Space Administration, the Japanese
    Monbukagakusho, the Max Planck Society, and the Higher Education Funding Council
    for England. The SDSS Web Site is http://www.sdss.org/. The SDSS is managed by the
    Astrophysical Research Consortium for the Participating Institutions. The Participating
    Institutions are the American Museum of Natural History, Astrophysical Institute Potsdam,
    University of Basel, University of Cambridge, Case Western Reserve University,
    University of Chicago, Drexel University, Fermilab, the Institute for Advanced Study, the
    Japan Participation Group, Johns Hopkins University, the Joint Institute for Nuclear Astrophysics,
    the Kavli Institute for Particle Astrophysics and Cosmology, the Korean Scientist Group,
    the Chinese Academy of Sciences (LAMOST), Los Alamos National Laboratory,
    the Max-Planck-Institute for Astronomy (MPIA), the Max-Planck-Institute for Astrophysics (MPA),
    New Mexico State University, Ohio State University, University of Pittsburgh,
    University of Portsmouth, Princeton University, the United States Naval Observatory,
    and the University of Washington. We also acknowledge the work of
the ALFALFA collaboration team in observing, reducing the data and constructing the catalogue of galaxies used in this
work.

%%%%%%%%%%%%%%%%%%%%%%%%%%%%%%%%%%%%%%%%%%%%%%%%%%

%%%%%%%%%%%%%%%%% APPENDICES %%%%%%%%%%%%%%%%%%%%%

%%%%%%%%%%%%%%%%%%%%%%%%%%%%%%%%%%%%%%%%%%%%%%%%%%

% Don't change these lines
\bsp	% typesetting comment
\label{lastpage}
\end{document}